\documentclass{mn2e}

\usepackage{epsfig}
 
\newcommand{\bff}{\textbf}

\newcommand{\noi}{\noindent}

\newcommand{\bc}{\begin{center}}
\newcommand{\ec}{\end{center}}
\newcommand{\bi}{\begin{itemize}}
\newcommand{\ei}{\end{itemize}}
\newcommand{\beq}{\begin{equation}}
\newcommand{\eeq}{\end{equation}}

\def \aa#1#2   {{A\&A} { #1}, {#2}}
\def \aas#1#2  {{A\&AS} {\bf #1}, {#2}}
\def \aj#1#2   {{AJ} { #1}, {#2}}
\def \apj#1#2  {{ApJ} { #1}, {#2}}
\def \apjs#1#2 {{ApJS} { #1}, {#2}}

\def \jcp#1#2   {{ J. Comp. Phys. \/} { #1}, {#2}}

\newcommand{\wt}{\widetilde}

\def \mb#1 {\mbox{\rm#1}}
\def \mbt#1 {\mbox{\tiny#1}}

\newcommand{\Msun}{\mbox{\,M$_{\odot}$}}

\newcommand{\cms}{\mbox{\,cm\,s$^{-1}$}}

\newcommand{\Msyr}{\Msun\,\mbox{yr$^{-1}$}}
\newcommand{\gcm}{\mbox{\,g\,cm$^{-3}$}}

\newcommand{\const}{\mbox{const}}
\newcommand{\rads}{\mbox{\,rad\,s$^{-1}$}}
\mathchardef\kap="001A
\def \half {\frac {1} {2}}

\def \ppder#1#2 {\frac {\partial #1} {\partial #2} }

\def \pp2der#1#2 {\frac {\partial^2 #1} {\partial #2 ^2} }

\def \nb#1 {{\it\bff{NB:}}{\it#1}}

\begin{document} 

\title[Stream--core interactions in massive star mergers]
{Hydrodynamical Simulations of the Stream--Core Interaction in the Slow
Merger of Massive Stars}
\author[N. Ivanova, Ph.\ Podsiadlowski and H.Spruit ]
{N.~Ivanova$^1$, Ph.~Podsiadlowski$^1$ and H.~Spruit$^2$ \\
$^1$ University of Oxford, Nuclear and Astrophysics Laboratory, 
Oxford, OX1 3RJ\\
$^2$ Max-Planck-Institut f\"{u}r Astrophysik, 
Karl-Schwarzschild-Strasse 1, 85741 Garching, Germany
}

\maketitle
 
\begin{abstract}
We  present  detailed  simulations  of  the interaction  of  a  stream
emanating  from a  mass-losing secondary  with the  core of  a massive
supergiant  in  the slow  merger  of the  two  stars  inside a  common
envelope. The dynamics  of the stream can be  divided into a ballistic
phase, starting at  the $L_1$ point, and a  hydrodynamical phase where
the stream  interacts strongly with the core.   Considering the merger
of a 1 and 5\Msun\ star with a 20\Msun\ evolved supergiant, we present
two-dimensional hydrodynamical  simulations using the  PROMETHEUS code
to demonstrate  how the  penetration depth and  post-impact conditions
depend  on the initial  properties of  stream material  (e.g. entropy,
angular  momentum,  stream  width)  and  the properties  of  the  core
(e.g. density  structure and rotation  rate). Using these  results, we
present  a   fitting  formula  for   the  entropy  generated   in  the
stream--core  interaction and a  recipe for  the determination  of the
penetration depth based on a modified Bernoulli integral.
\end{abstract}

\begin{keywords}
binaries : close ---  hydrodynamics
--- nucleosynthesis
\end{keywords}
 
\section{Introduction}

A large  fraction of massive stars  are known to be  members of binary
systems  which are close  enough to  interact strongly  via Roche-lobe
overflow (see  e.g. Garmany, Conti \&  Massey 1980).  One  of the most
dramatic, but poorly understood  types of binary interactions involves
the complete merger  of the two components in  a common envelope which
leads  to  the formation  of  a  single,  rapidly rotating  star  with
possibly  some  highly  unusual  properties. This  type  of  evolution
generally occurs as the result of dynamical mass transfer (Paczy\'nski
\&  Sienkiewicz 1972)  where a  more  massive star  (usually a  giant)
overfills its Roche lobe by  an ever increasing amount leading to mass
transfer on a  dynamical time-scale. Since the companion  is only able
to accrete matter on its thermal time-scale, most of the excess matter
is believed  to form a common  envelope engulfing both  the mass donor
and the secondary. Inside this envelope, the core of the giant and the
secondary form a close  binary.  Friction between this immersed binary
and  the envelope  then  causes the  orbit  to shrink.  If the  energy
released  in the  process is  sufficient to  eject the  envelope, this
leads to the  formation of a very close binary  consisting of the core
of  the giant and  a largely  unperturbed companion  star (Paczy\'nski
1976; Taam \& Sandqvist 2000). However, if the envelope is not ejected
in the early rapid spiral-in phase,  this phase is followed by a phase
of self-regulated spiral-in where  all the frictional energy deposited
in the  envelope can be transported  to the surface where  it is being
radiated   away  (Meyer   \&   Meyer-Hofmeister  1979;   Podsiadlowski
2001). For a massive primary with a mass of $\sim 20\Msun$ filling its
Roche lobe as  a red supergiant after the end  of helium core burning,
the characteristic  time-scale for the  self-regulated spiral-in phase
is several 100\,yr. However, at some point, the spiraling-in secondary
itself will fill its critical tidal lobe and start to transfer mass to
the core of  the giant. This situation is in  many respects similar to
the case  of normal  mass transfer by  Roche-lobe overflow  except for
several  fundamental differences.  (1) Mass  transfer occurs  inside a
low-density, opaque  envelope, where the density  contrast between the
envelope  and  the secondary's  average  density  may  be as  high  as
1:$10^6$.   (2) The  mechanism for  driving mass  transfer  is orbital
angular momentum loss due to  the friction of the immersed binary with
the envelope. Because of the  short spiral-in time-scale this leads to
extremely high  mass transfer  rates of order  $0.01 -  10\Msyr$.  (3)
Since  the  spiral-in  time-scale  is  much  shorter  than  a  typical
synchronization time-scale  in a radiative star (e.g.  Zahn 1975), the
spin of the secondary may not be synchronized with the orbit.

As  a result  of this  interaction, the  secondary is  gradually being
dissolved inside  the common envelope. Even though  the time-scale for
this dissolution is short compared to the evolutionary time-scale of a
giant, it  is still much longer  than the dynamical  time-scale of the
secondary.  We therefore  refer to this process as  a `slow merger' of
the components,  although this phase is  likely to be  terminated by a
delayed  dynamical  instability  when  the flat-entropy  core  of  the
secondary    is   being    exposed   and    mass    transfer   becomes
catastrophic. This will ultimately lead to the dynamical disruption of
the remnant core (if it is relatively unevolved).

In this paper  we present the first part of a  systematic study of the
slow merger of massive stars,  concentrating on the interaction of the
mass stream emanating from the mass-losing immersed star with the core
of the giant. This stream will generally not intersect with itself and
form an accretion disc, but  will instead penetrate deep into the core
of the  giant before it is  stopped by the  increasing pressure inside
the core. As we will show, the stream may be able to reach a radius as
small  as   $\sim  10^{10}\,$cm.  Since   this  is  deep   inside  the
hydrogen-exhausted  helium core of  a 20\Msun\  star that  has already
exhausted helium  in the core,  this may have  important consequences:
(1) It can  lead to the  dredge-up of helium  from the core  which can
then be mixed with the rest of the envelope. (2) Since the material in
the stream is  hydrogen-rich and is heated to  temperatures as high as
several $10^8\,$K  in the  post-impact region, this  may lead  to some
unusual  nucleosynthesis,  in particular  s-processing.  How deep  the
stream  can penetrate  depends on  the initial  entropy of  the stream
material, but  also its initial angular momentum  and most importantly
on the structure near the core  and the amount of entropy generated in
the interaction  with the core.   It is the  purpose of this  paper to
systematically explore  how the  penetration depends on  these various
factors   and  to   determine  the   properties  of   the  post-impact
material. In  a subsequent paper, we  will use these  results to model
the slow merger of two massive  stars in the context of the progenitor
of SN~1987A,  whose highly anomalous properties are  likely the result
of  such a  merger (see  Podsiadlowski 1992,  1997 for  discussion and
references).  While the hydrodynamical  simulations presented  in this
paper were  designed with  this application in  mind, our  analysis is
sufficiently general  to allow application to other  systems that have
been suggested  to be  post-merger objects (e.g.  V Hydrae  [Kahane et
al.\  1996]; FK  Comae  stars  [e.g. Rucinski  1990;  Welty \&  Ramsey
1994]).

In  Section~\ref{streaml1}  we   outline  the  basic  assumptions  for
modelling the initial properties of the stream leaving the mass-losing
secondary.  In  Section~\ref{streambal} we use this model  to obtain a
description  of  the  stream   at  the  start  of  the  hydrodynamical
simulations.   Section~\ref{method}  provides  a  description  of  the
numerical  method   used  in  the   hydrodynamical  calculations.   In
Section~\ref{streamhydro}   we  present   the   results  of   detailed
hydrodynamical simulations  and systematically discuss  the physics of
the interaction between the jet  and the ambient matter, in particular
the mechanism of  the stream dissipation, the effects  of rotation and
of nuclear burning in the stream.  Finally, in Section~6 we present an
semi-empirical prescription  which allows to  estimate the penetration
depth, taking into account the entropy generated in the interaction.

\section{The initial conditions of the stream} 
\label{streaml1}

\subsection{Mass-loss rates}

Once the immersed secondary starts to fill its tidal lobe, it starts to
transfer mass to the core of the giant. The mass-transfer rate is determined
by the friction between the spiraling-in immersed binary and the envelope
which drives the evolution of the system. Since the mass-loss time-scale
is generally much shorter than the nuclear or thermal time-scale for
a main-sequence star, the radius responds adiabatically to mass loss, i.e.

\beq
{\frac {\dot R_{\rm 2 }} {R_{\rm 2 }}} = 
\left ( \ppder {\ln R_{2}}{ \ln M_{\rm 2 } } \right )_{\rm ad }
{\frac {\dot M_{\rm 2 }} {M_{\rm 2 }} } = 
\zeta_{\rm ad } {\frac {\dot M_{\rm 2 }} {M_{\rm 2 } } }, \nonumber
\eeq
where $R_2$ and $M_2$ are the radius and the mass of the mass-losing
secondary, respectively.

\noi The Roche-lobe radius $R_{\rm RL}$ of the secondary changes according to

\begin{eqnarray}
{\frac {\dot R_{\rm RL }} {R_{\rm RL }}} &=&  
\left ( \ppder {\ln R_{\rm RL }}{ \ln M_{2} } \right )_{D=\rm const}
 {\frac {\dot M_{\rm 2 }} {M_{\rm 2 } } }
+ \left( \ppder {\ln R_{\rm RL }}{t} \right )_{\dot M_{\rm 2 } = 0} \nonumber \\
&\equiv&\zeta_{\rm RL } {\frac {\dot M_{\rm 2 }} {M_{\rm 2 } } }
+ \left( \ppder {\ln D}{t} \right )_{\dot M_{\rm 2 } = 0},
\end{eqnarray}

\noi where the Roche-lobe radius $R_{\rm RL }$ is given by Eggleton (1983)

\beq
\ R_{\rm RL }  = {\frac {0.49\, q^{2/3} } 
{0.6\, q^{2/3}+\ln(1+q^{1/3}) } }\, D.\ \nonumber
\eeq

\noi Here $q=M_2/M_1$ is the mass ratio, and $M_1$ is the total mass of 
the primary core within the binary separation $D$. 

Assuming that the radius of the star is equal to the Roche-lobe radius,
we can equate equations (1) and (2) and obtain an estimate for the mass-loss
rate from the secondary as

\beq
  {\frac {\dot M_{\rm 2 }}
{M_{\rm 2 }} }
= {\frac {1} {\zeta_{\rm ad }-\zeta_{\rm RL }} } 
\left ( \ppder {\ln D}{t} \right )_{\dot M_{\rm 2 }  = 0}  , \nonumber
\eeq

\noi where $\partial \ln D/\partial t$ is the change in the orbital separation
due to tidal and viscous friction. For typical conditions in a 20\Msun\
supergiant, the resulting mass-loss rates are typically in the range
of 0.01 to 10\Msyr (Ivanova, Podsiadlowski \& Spruit 20001; and
Ivanova \& Podsiadlowski 2002 [IP]). 
To obtain values for $\zeta_{\rm ad }$,
we performed separate stellar-evolution calculations where mass was
taken off the assumed secondary on a time-scale much shorter than the
thermal time-scale in its outer envelope
(see e.g. Podsiadlowski, Rappaport \& Pfahl 2001).

\subsection{The stream density profile and the stream width}

In previous treatments of mass outflow from $L_1$, it is generally assumed
that the stream is isothermal
(see e.g. Lubow \& Shu 1975; LS;  Bisikalo et al.\ 1997).  
Because of the high expected mass-loss rate  in our problem, it is more
appropriate to assume that the stream behaves adiabatically, since the
radiative diffusion time-scale is generally much longer than the
characteristic flow time-scale (for a stream with a radius of
$10^{10}\,$cm, the diffusion time is of order
$10^6\,$s). 

In the case where the secondary's spin is aligned and synchronous with
the orbit, we can determine the density distribution in the neighbourhood
of the $L_1$ point by expanding the potential around this point 
in a Taylor series (see e.g. LS), 
where we use Cartesian coordinates $(x,y,z)$:

\begin{eqnarray}
\label{phi_taylor}
\Phi &=& \Phi_{L1} -\left ( 2 A + 1   \right  )   
{\frac {(x-x_{L1})^2} {2}} \Omega^2  + \nonumber \\
&&\hspace{2truecm}
 \left ( A -1 \right )  {\frac {y^2} {2}} \Omega^2 
+ A {\frac {z^2} {2}} \Omega^2
\  , \nonumber \\
A &=& {\frac {\mu} {|x_{L1}/D-1+\mu|^3} } +{\frac {1-\mu} {|x_{L1}/D-\mu|^3}.}
\end{eqnarray}

\noi Here $\Omega$ is the angular speed of the binary and 
$\mu =  M_{\mb 1 } / (M_{\mb 1 } + M_{\mb 2 } )$.

A natural assumption is that the energy distribution within the stream,
consisting of kinetic and internal energies, follows the 
shape of the Roche potential and that the Bernoulli integral 
for the matter leaving the donor star is conserved (for an adiabatic
stream)

\beq 
B \equiv \Phi + {\frac  {u^2} {2}  } +  {\frac {c^2}  {\gamma-1}} =
\Phi_{L_1} + {\frac  {u_{{L_1}^2}} {2}  } +  {\frac {c_{L_1}^2}  {\gamma-1}}
\ .
\label{Bern_int}
\eeq

\noi Here $u_{L_1}$ and $c_{L_1}$ are the velocity and the sound speed
of the matter at $L_1$, and $\gamma$ is the adiabatic index of the stream
material. Let us assume that in the $L_1$ neighbourhood
the stream is laterally in hydrostatic equilibrium. 
This is a good approximation, especially in the $Z-$direction
(LS).  Then, in the $YZ-$plane 

\beq
 \Phi - \Phi_{L1} = 
\left ( {\frac {A-1} {2}} y^2 + {\frac {A} {2}} z^2 \right ) 
\Omega^2 \approx {\frac {\alpha} {2}} r^2 \Omega^2,
\label{c_l1}
\eeq

\noi where $r$ is the distance to the $L_1$ point and $\alpha \approx A$
is a unitless constant which depends on the 
shape of the potential in the neighbourhood of $L_1$.
For binary systems with a  mass ratio, $q$, between 0.006 and 1.0, it has 
a value of 6.16 to 8.0 (LS).
Then the density distribution within an adiabatic stream
can be obtained from the condition of lateral hydrostatic equilibrium as

\beq
\rho  = \rho_{\rm c}  \left ( 1- {\frac {\alpha} {2}} {\frac {r^2} {\wt W^2}} 
\right ) ^ {1/(\gamma-1)}
\label{eq:dens}
\eeq

\noi where $\rho_{\rm c}$ is the density of the material 
in the core of the stream, and
$\wt W$ is the characteristic width of the stream, which
depends on the properties (e.g. entropy, temperature) of the matter
in the secondary:

\beq
\wt W = \sqrt{\frac {\gamma } {\gamma-1}} 
\frac { s^{\half} \rho_{\rm c }^{1/3}} { \Omega} .
\eeq

\noi  
Here $s=P/\rho^\gamma$ is the adiabatic constant of the material from
the secondary.  
An adiabatic stream can not have a width wider than 
$W = \wt W\sqrt{2/\alpha}$ (as can be seen from equ.~\ref{eq:dens}).

We assume that
the velocity of the material at $L_1$ is given by the local sound
speed (see LS) and that this speed varies through the stream cross-section.
Then, integrating the mass-flow rate across the stream cross section near
$L_1$, we can relate the mass-loss rate  $\dot M_{\rm 2}$ (in units
of \Msyr) to the density in the stream core for an adiabatic stream (with 
$\gamma=5/3$):

\begin{eqnarray}
\rho_{\rm c} &=& 1.4 \times 10^{-3}\mbox{g\,cm}^{-3}  \\
&&\hspace{1cm} \dot M_{\rm 2}^{1/2} \Omega_4 s_{15}^{-3/4} 
\left (  x_{\rm s }^2 - {\frac {\alpha} {2}} x_{\rm s }^4 + 
{\frac {\alpha^2} {12}} x_{\rm s }^6
\right ) ^ {-\half},\nonumber
\label{ml_ad}
\end{eqnarray}

\noi where  $s_{15} = ({P} / {\rho^{\gamma}})/(10^{15}\mbox{cgs})$ 
is the scaled adiabatic constant of the material from the secondary,
 $\Omega_{4} = \Omega / 10^{-4}\mbox{s}^{-1}$,
$x_{\rm s } = W/\wt W$ is the unitless stream width. 

When the stream forms in the hot and high-pressure environment
of a common envelope, the actual stream width is determined
by the local conditions of the envelope.
If we assume that the outer edge of the stream is in pressure 
balance with the common envelope, the stream density at the edge
of the stream cannot be less than

\beq
\rho_{\rm out } =  
\left (P_{\rm CE}/s \right ) ^{1/\gamma} \ ,
\eeq
 
\noi where $P_{\rm CE}$ is the ambient pressure in the common envelope.
The stream width for the current mass loss rate 
then can be found from the equ.~(10) in combination with

\beq
\rho_{\rm c} = 
\left (P_{\rm CE}/s \right ) ^{1/\gamma}  
\left ( 1- {\frac {\alpha} {2}} x_{\rm s }^2 
  \right ) ^ {-1/(\gamma-1)} \ 
\eeq
\noi (from equ.~8).

\subsection{The effect of non-synchronous rotation}

Since the secondary's spin may not be synchronous with the orbit (in
fact, it may be rotating faster or more slowly depending on the
circumstances), we also need to consider the case of non-synchronous
rotation.  For this purpose we adopt a coordinate system that rotates
with the secondary.  At the equipotential surface which determines
the surface of the secondary (the donor), there is an area where matter is
being accelerated outwards from the stellar surface.  From this zone
matter may flow out.  In the case of a synchronized secondary, the
physical situation is a close analogue of a Laval nozzle, 
where the equipotential surfaces take over
the role of a nozzle.  In the standard case, when a thin flow is
considered, $L_1$ can be taken as a point.  Then, as for the case of
the standard Laval nozzle, matter is almost completely unaffected by
external acceleration at the nozzle mouth (at the sonic point), but
experiences negative acceleration before and positive acceleration
after passing though the nozzle.

For an asynchronously rotating secondary, the position
of the positive acceleration zone with respect to the secondary
changes with time. To parametrize the degree of a-synchronicity, we
introduce a parameter $f$,

\beq
f = {\frac {\Omega_{\mbox {\scriptsize sec} }} {\Omega} } - 1,
\eeq

\noi where $\Omega_{\rm {sec} }$ is the angular velocity of the secondary. 
This parameter is zero if the secondary rotates synchronously.
The position of the point  where the acceleration experienced by
matter on the stellar surface becomes non-negative can be found similarly
to the $L_1$ point (in order to avoid confusion with the standard $L_1$
point, we refer to this point as the `leaving point', $LP$),
considering the generalised Roche potential, $\Psi$, for the frame with the
origin at the center of the secondary, which rotates with
the secondary (see e.g. Limber 1963): 

\beq
\label{as_pot}
\Psi = \Phi - {\frac {\zeta^2 + \eta^2} {2} } f (f + 2) \Omega^2 \ .
\eeq

\noi Here $\Phi$ is the usual potential of the binary

\beq 
\Phi = -{\frac {G M_1} {|{\bmath r_1}|} } -{\frac {G M_2} 
{|{\bmath r_1}|}} - {\frac {1} {2}}
\left ( \Omega \,{\bmath e_z} \times {\bmath r} \right ) ^2 \ 
\label{eq:full_potential_corot}
\eeq

\noi and ($\zeta$,
$\eta$) are Cartesian coordinates in the frame rotating with the
secondary.  For a secondary which rotates faster than the binary
itself, $LP$ will normally be shifted from $L_1$ towards the donor;
and for a secondary which rotates more slowly than the binary this point
will be shifted towards the primary.  A rough estimate of the linear
speed with which the zone of positive acceleration travels across the
secondary surface, $v_{\mbox {\scriptsize zone} } = f\Omega_{\rm sec } 
r_{\rm sec }$, shows that it is typically
twice the local sound speed of the secondary (for the case of $f=1$
and considering parameters characteristic for a system at the point 
where the secondary starts to overfill its Roche lobe).  
In this case, we do not have a
normal nozzle (as for a synchronized secondary), as matter at a
particular position on the secondary's surface (in the frame rotating
with the secondary) is not affected by any matter that has escaped
from the secondary before it enters the zone of non-negative
acceleration.  Therefore we may consider what happens to a fluid
element on the secondary independently from any matter that has flowed
out previously. A physical analogue of this situation is a ball,
containing the (secondary) matter, that is separated from the
outside by negative acceleration.  At the surface of the ball there is
a hole which travels with supersonic speed.  As far as the matter in
the ball is concerned, the surface of the ball acts like a wall that
is suddenly being removed (for some period of time) and then placed
back again.  Matter will escape from the ball through the hole only if
the duration of the interval of positive acceleration is long enough
for matter to be able to escape before it is pulled back by the
negative acceleration.  Another analogue of the problem is the sudden
expansion of a gas sphere into a vacuum. In this case, the speed with
which matter expands can be as high as 3 times the sonic speed (see
e.g. Zeldovich \& Raizer 1966).
In a real situation, we may expect a combination of
these and other effects that cannot be easily treated analytically.
As this discussion shows, for an a-synchronously rotating
secondary, the range of the possible initial flow velocities includes
significantly supersonic speeds. Fortunately, our ballistic calculations
in \S~3 show that the stream properties close to the primary core
are not strongly dependent on the initial flow velocity in the stream.

\subsection{The escape angle - generalization of the LS formalism}

In the case where the secondary's spin is not synchronous with the
orbit, the angle at which matter leaves the secondary will be different
from the synchronous case analyzed by LS.
To obtain this angle in our problem, we generally follow the LS formalism
with the following changes: we place the origin of the 
coordinate system not at $L_1$  but at LP, and we use a coordinate frame 
which rotates with the secondary. 
Using cylindrical coordinates, the potential in the neighbourhood 
of $LP$ can be expanded (in the $z=0$ plane) as

\beq
\Phi = \Phi_{LP} - \left [  
1 + f(f+2) + {\frac {A_{LP}} {2}} (3 \cos 2 \theta +1 ) 
\right ] 
{\frac {r^2} {2} } \Omega^2,  
\eeq

\beq
A_{LP} = {\frac {\mu} {|x_{LP}/D-1+\mu|^3} } + {\frac {1-\mu} {|x_{LP}/D+\mu|^3} },
\eeq

\noi where $X_{LP}$ is the distance between the center of mass and $LP$
and $\theta$ is the angle between the flow direction of the stream
and the direction of the primary.

The escape angle of the stream in the frame rotating with 
the secondary can then be found, analogously to the derivation
in LS, from the equation

\begin{eqnarray}
\lefteqn{  \cos^2 2 \theta_s + {\frac {8} {3}} \cos  2 \theta_s }
\nonumber \\
&+& {\frac {16} {9 A_{LP}^2} }
\left ( 
{\frac {A_{LP}} {2} } + 1 + f(f+2) - {\frac {9 A_{LP}^2} {16} }
\right ) = 0.
\end{eqnarray}

\noi For the case of a synchronously rotating secondary (f=0), this 
equation reduces to the solution for $\theta_s$ obtained by LS. 
The angle with which the stream will escape from $LP$ in
the frame co-rotating with the binary is given by the transformation

\beq
\tan \theta = {\frac { f \Omega  D (1 - \mu + x_{LP}/D) + v_e \sin \theta_s}  
{v_e \cos \theta_s} },
\eeq

\noi where $v_e$ is the velocity of the stream at the nozzle 
(sonic for a synchronously rotating secondary).

\section{The ballistic phase} \label{streambal}

The motion of the stream can be naturally divided into two phases, a
(quasi-)ballistic phase where the stream trajectory is not strongly
affected by the interaction with the ambient matter, which can be
treated more or less ballistically, and a hydrodynamical phase where
the dynamics of the stream is dominated by the hydrodynamical
interaction with the core.  Considering these two phases separately
helps to avoid numerical problems (due to the complexities near the
$L_1$ point) at the start of the hydrodynamical calculation. However, it
also introduces some uncertainty since matter in the inner region
surrounding the primary core may not be in complete co-rotation with
the orbit and will have a backward motion with respect to the
co-rotating frame.  Such relative rotation causes a force which
reduces the interception angle between the falling stream and the
normal direction to the core relative to the predictions from a purely
ballistic trajectory (see also \S~5.3.1).

The equations of motion of the stream are

\begin{eqnarray}
\lefteqn{ u_x \ppder {u_x}{x} + u_y \ppder {u_x}{y} = - \ppder {\Phi}{x} + 
2u_y  \Omega- 
{\frac {1} {\rho}} \ppder {P}{\boldmath l } {\frac {u_x} {u}}  \ ,}
\nonumber \\
\lefteqn{  u_x \ppder {u_y}{x} + u_y \ppder {u_y}{y} = - \ppder {\Phi}{y} - 2u_x\Omega - 
{\frac {1} {\rho}} \ppder {P}{\boldmath l } {\frac {u_y} {u} }\ .}
\end{eqnarray}

\noi Here $u=\sqrt{u_x^2+u_y^2}$ is the magnitude of the central stream
velocity, and $u_x, u_y$ are its components. 
The pressure gradient along the path $  (1/\rho)\, \partial{P}/\partial{l}$ 
is included for completeness. 
This term is generally small, since the Mach number of the    
flow is large except near the nozzle. It is kept here for completeness and  
consistency in the calculations.
The structure of the common envelope enters into these
equations through the pressure boundary condition at the stream edge, which
is set equal to the ambient pressure in the envelope.
The density within the stream is related to the pressure 
by  the adiabatic constant of the secondary material 
$s = P/\rho^\gamma = \const$.  
Together with a specified mass-flow rate, this set of equations provides
a complete set that completely describes the dynamics of the
stream in this phase.

\subsection{Initial conditions for the hydrodynamical calculations}

The initial conditions for the hydrodynamical simulations are taken
directly from the parameters obtained from the ballistic calculations
for a range of binary parameters.  Specifically, we consider
parameters representing the spiral-in and subsequent merger of a
secondary of 1 and 5\Msun\ inside a 20\Msun\ evolved supergiant which
has completed helium burning in the core and has a core mass of $\sim
7\Msun$. To model the secondaries, we first followed their evolution
to the same age as the primary and then evolved them further,
subjecting them to very fast mass loss to model their adiabatic
response to mass loss. These calculations give the radii of the stars
and their surface entropies as a function of current stellar
mass. 

\section{The numerical method} \label{method}

\subsection{The hydrodynamics code} 

For the hydrodynamical simulations of the stream--core
interactions we use a code based on the PROMETHEUS  hydrodynamical code 
(Fryxell, M\"{u}ller \& Arnett 1989).
This is an Eulerian code, which uses a second-order Godunov-type scheme
to solve the hydrodynamical equations,
the  piecewise parabolic method (PPM) of Colella \& Woodward (1994).
This code has been widely used by different groups (see e.g.
Kercek, Hillebrandt \& Truran 1998; KHT), and we refer to these papers
for a detailed description of the code and numerical tests.
To make it applicable to our problem, we had to make a number of  modifications 
to the original code, in particular to the treatment
of the gravity field, the boundary conditions, 
to the equation of state and the hydrodynamical equations themselves
(to take into account the effects of a frame co-rotating with the binary).
Since the mass transfer occurs  in an opaque environment, radiative
losses are not important and have been neglected.

In the presence of a large pressure  gradient at the core--envelope 
interface and a correspondingly strong gravitational field,
Godunov-type schemes produce intrinsic 
accelerations, which during a few dynamical time-scales $t_{\rm d } $
create significant outward motion (KHT).
For a first-order Godunov method it is possible to find an
analytical formula to modify the interface states. Then source terms
(the gravitational field in our case) will be balanced by flux differences
(LeVeque 1998). 
This is not possible for higher-order Godunov-type schemes, where
each problem requires a separate treatment (KHT).
To reduce the artificial acceleration in our case, 
we assume that the gravitational field can be considered as constant
in time and that there is no self-gravity.
Then, in each time step, we modify the interface values before 
applying the Riemann solver, by reducing them 
from both (left and right) interfaces by a
pressure flux caused by gravity. Furthermore, we enforce the condition 
of hydrostatic equilibrium at the boundary 
(for ghost cells\footnote{Ghost cells are a few additional cells (four
in the case of a PPM code) at each boundary of the computational domain,
where the values at these cells are determined by the boundary 
conditions.}) in the direction of the gravity field.
These modifications ensure that any initial model
remains stable for arbitrarily long time.

\subsection{The model set-up}

We performed 2-dimensional calculations in polar cylindrical
coordinates, where we usually used $300 \times 300$ grid points. In
the case of very narrow streams, we increased the resolution to $600
\times 600$ grid points.  We also carried out test calculations with
grids of higher resolution to satisfy ourselves that further
increasing the number of grid points does not significantly 
affect the results.  As
a general rule, we ensure that the inflowing stream contains at
least 20 grid points in the azimuthal direction. We also carried out 
a few 3-dimensional calculations (in spherical coordinates) 
for the case of a symmetrical 
(non-inclined) stream in a non-rotating frame.
The results of these calculations showed that there were no significant
differences between the 2- and 3-dimensional calculations with the
same initial parameters. In particular, the penetration depths of
the stream were very similar in both cases.

\subsubsection{Ambient matter}
To model the region of interest for the stream--core interaction
inside the common envelope, we adopt power-law distributions for the
temperature and the pressure, i.e $ T(r)=T(r_0) \left (
{{r_0}/{r}}\right )^{\alpha_{T}} $ and $P(r)=P(r_0) \left
({{r_0}/{r}}\right )^{\alpha_{P}}.$ We fitted these power laws to the
structure in actual CE calculations (as described in detail in IP).
Typical values for $\alpha_{T}$ and $\alpha_{P}$ are
in the range $(1.3;5.2)\div (0.8;3.2)$, but can be as high
as $(1.7;7)$ 
(corresponding to a structure with an adiabatic index 
$\gamma_{\rm amb } \approx1.44$) and reasonably describe the
regions of the stellar models at the the evolutionary stage of
interest (IP).  For our calculations here, we adopted
parameters $ (\alpha_{ T } ; \alpha_{ P } ) = (1.3 ; 5.2)$ and $(0.8;
3.2)$ for models representing a 20+1 and a 20+5 CE simulation,
respectively.
Throughout the domain of our calculations, we assume that the ambient
matter has a constant composition, similar to the composition in
the core region of the primary, mainly helium. In the full stellar models,
hydrogen is exhausted below a radius of $4\times 10^{10}\,$cm, and the
hydrogen mass fraction increases to $\sim 0.1$ at the outer edge
of the domain (at $7.5\times 10^{10}\,$cm ).
We have tested that  these differences
do not affect the results appreciably. 
In this parametrized model, we assume that the ambient matter
is initially in (quasi-)hydrostatic equilibrium; 
as the model is parametrized by the pressure and temperature gradient,
the gravitational field is then defined implicitly
by the initial pressure gradient of the frame and 
the initial density distribution. 
In calculations where nuclear burning is included,
we use a nuclear reactions network with 27 isotopes, which includes
all important reactions up to $^{25}$Mg, in particular all $\alpha$ reactions
and the complete hot ($\beta$-limited) CNO cycle.
The reaction rates were taken from 
the  Thielemann nuclear reactions library (provided by R. C. Cannon)
The changes in chemical composition and the energy density due to the
nuclear burning are updated after each time step in the hydrodynamical
calculation. To prevent nuclear burning in the ambient medium, we
assumed that the ambient medium contained neither H nor He in these
particular calculations.
(For further details we refer to IP.)

\subsubsection{Boundary conditions}
The boundary conditions in the code, which describe the characteristics
of the inflowing stream, are the internal and external Mach numbers, 
$M_{\rm int}$ and $M_{\rm ext}$; 
the ratio between the central density in the stream and 
the ambient density at the outer boundary 
$\eta_{\rho}$; the initial inclination angle
of the stream $\theta$ (i.e. the angle between the initial flow direction and
the direction of the primary). For the chemical composition of the gas 
in the stream we use the abundances of the secondary.
 
The boundary conditions for the rest of the box in the radial
direction are outflow boundaries  (except in the gas inflow zone) where
the condition of hydrostatic equilibrium has been imposed for ghost cells.
The boundary condition in the
azimuthal direction also assumes outflow conditions (we used a grid which
normally covers an angle $\pi$). 
As  Courant number we used  $N_{\rm CFL } = 0.6 \div 0.8$.
A further parameter is the angular velocity with which the coordinate
system rotates (most calculations were performed in
the frame rotating with the primary core).

\section{Interaction between the stream and the ambient matter} \label{streamhydro}

\subsection{The initial stream-core interaction: establishing a steady stream}

\begin{figure*}
\centering
\includegraphics[scale=0.45]
 	{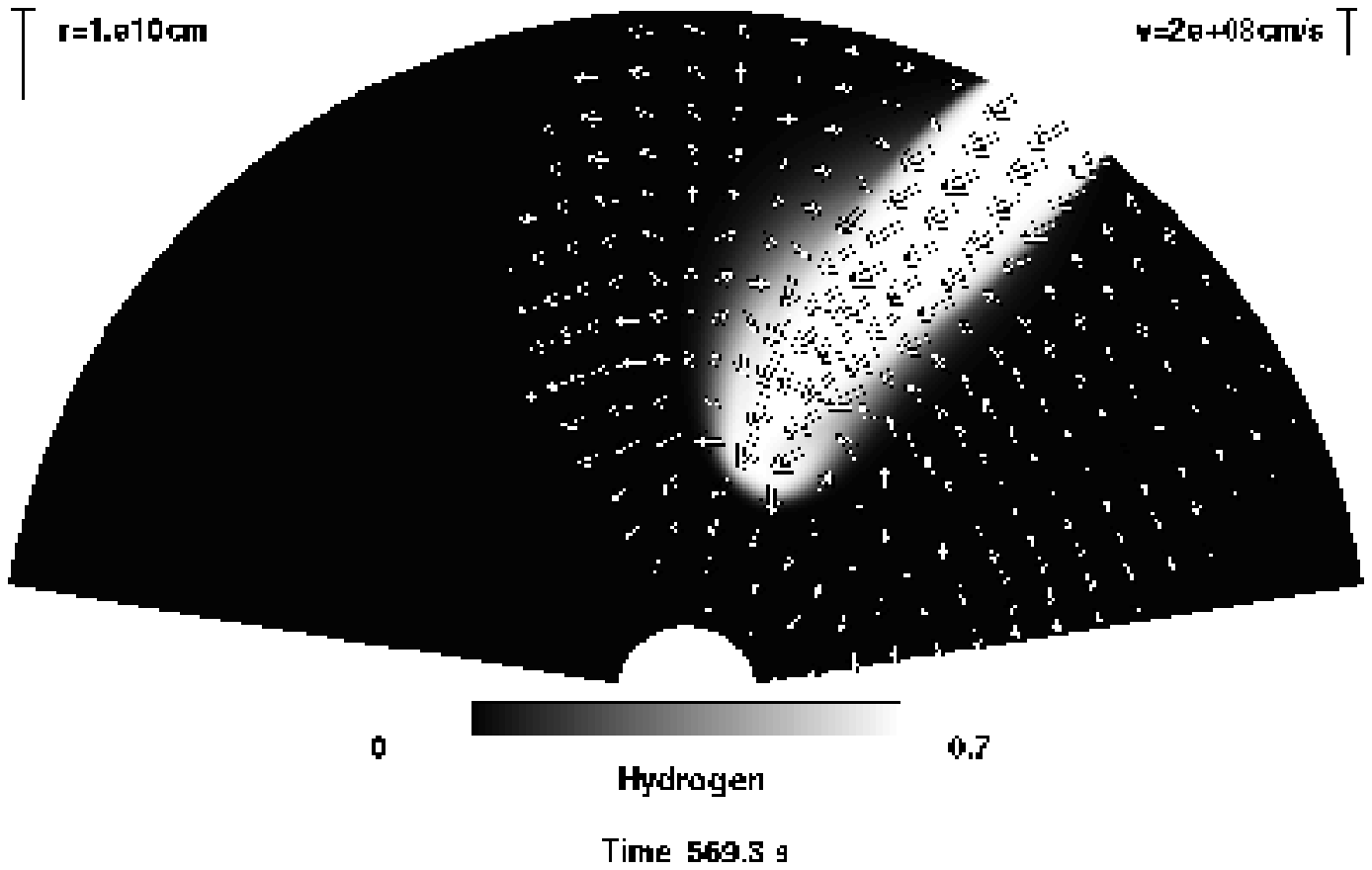}\hspace{2cm}
\includegraphics[scale=0.45]
 	{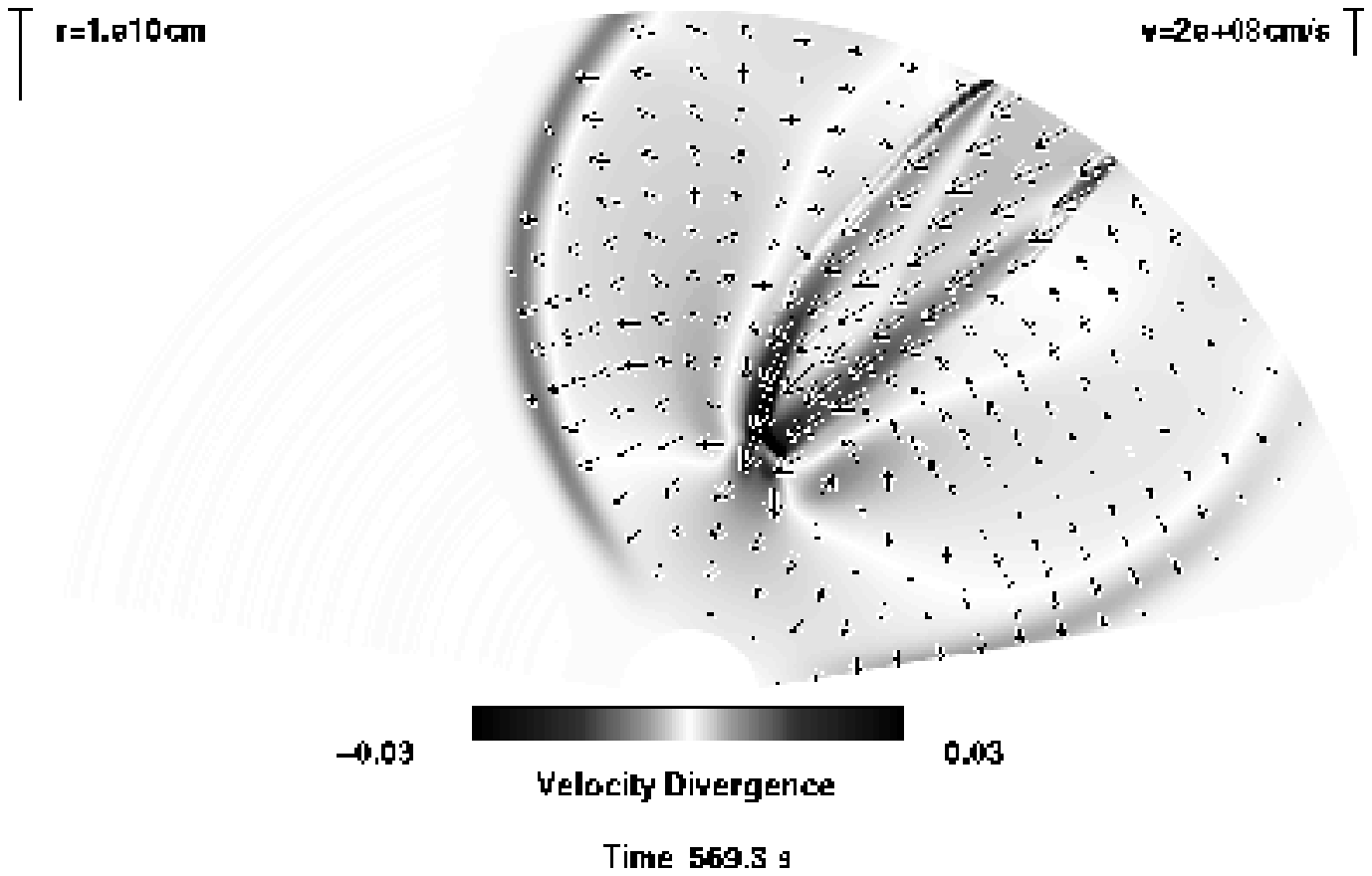} \\
\includegraphics[scale=0.45]
 	{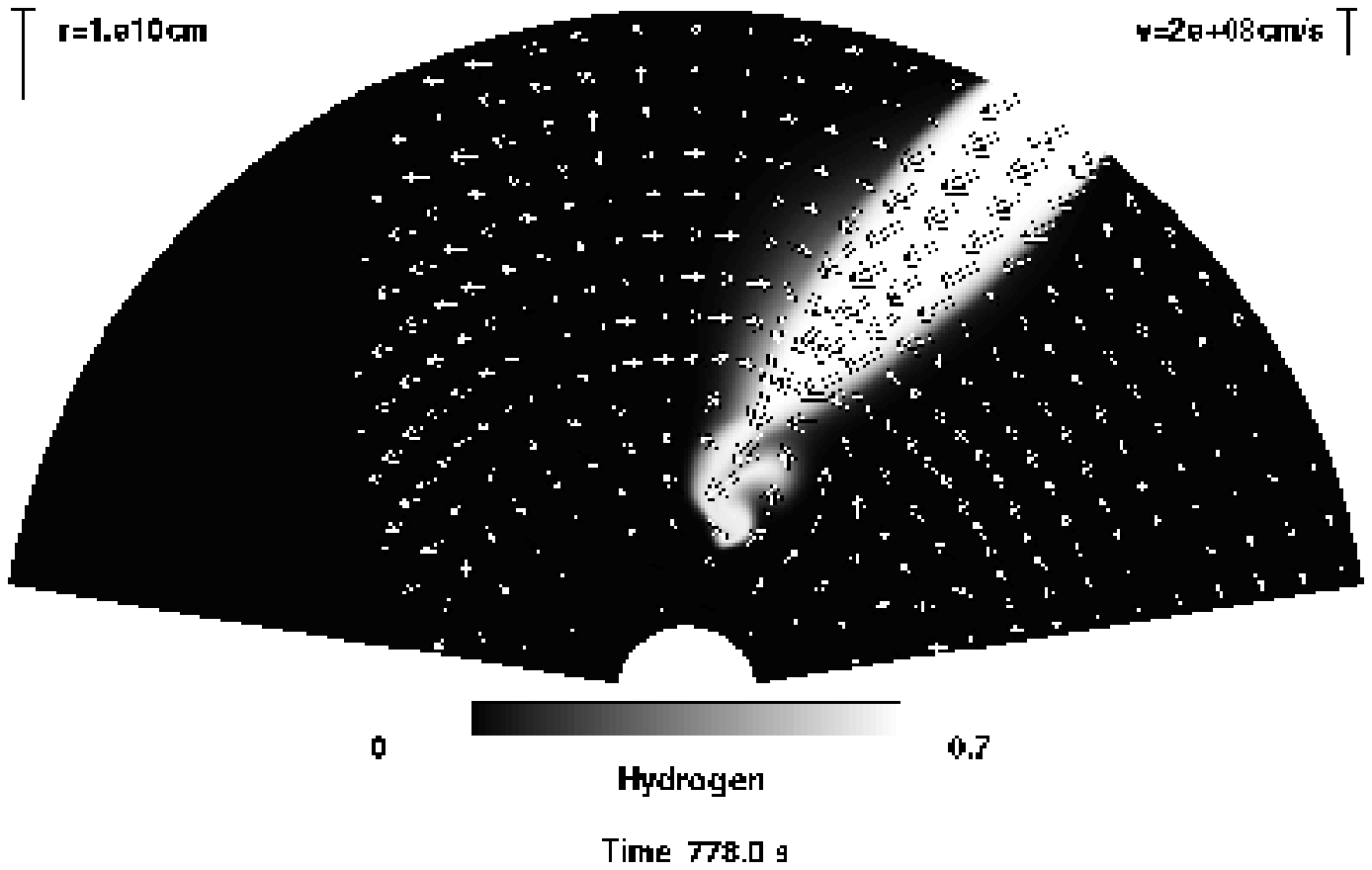}\hspace{2cm}
\includegraphics[scale=0.45]
 	{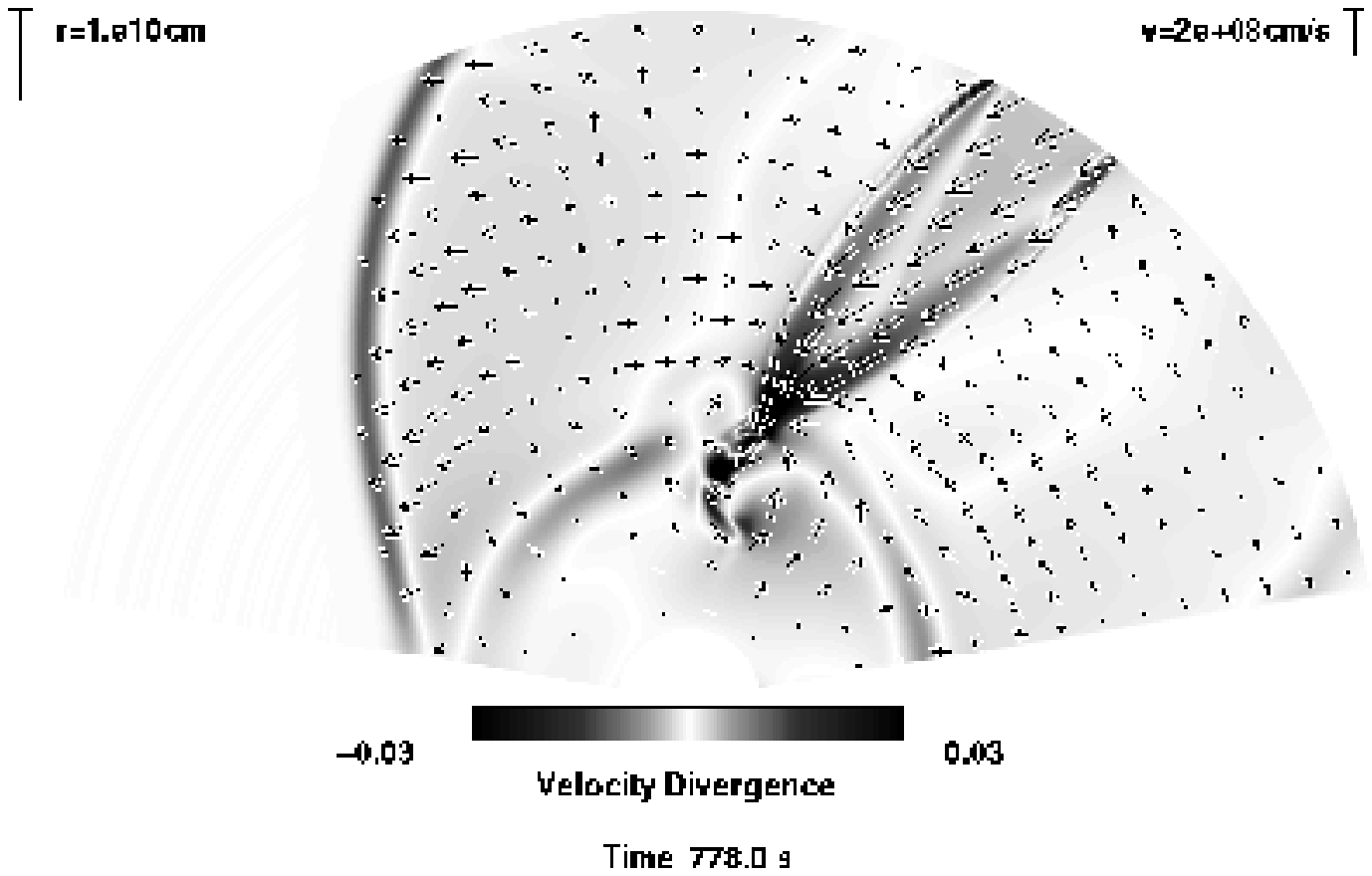} \\
\includegraphics[scale=0.45]
 	{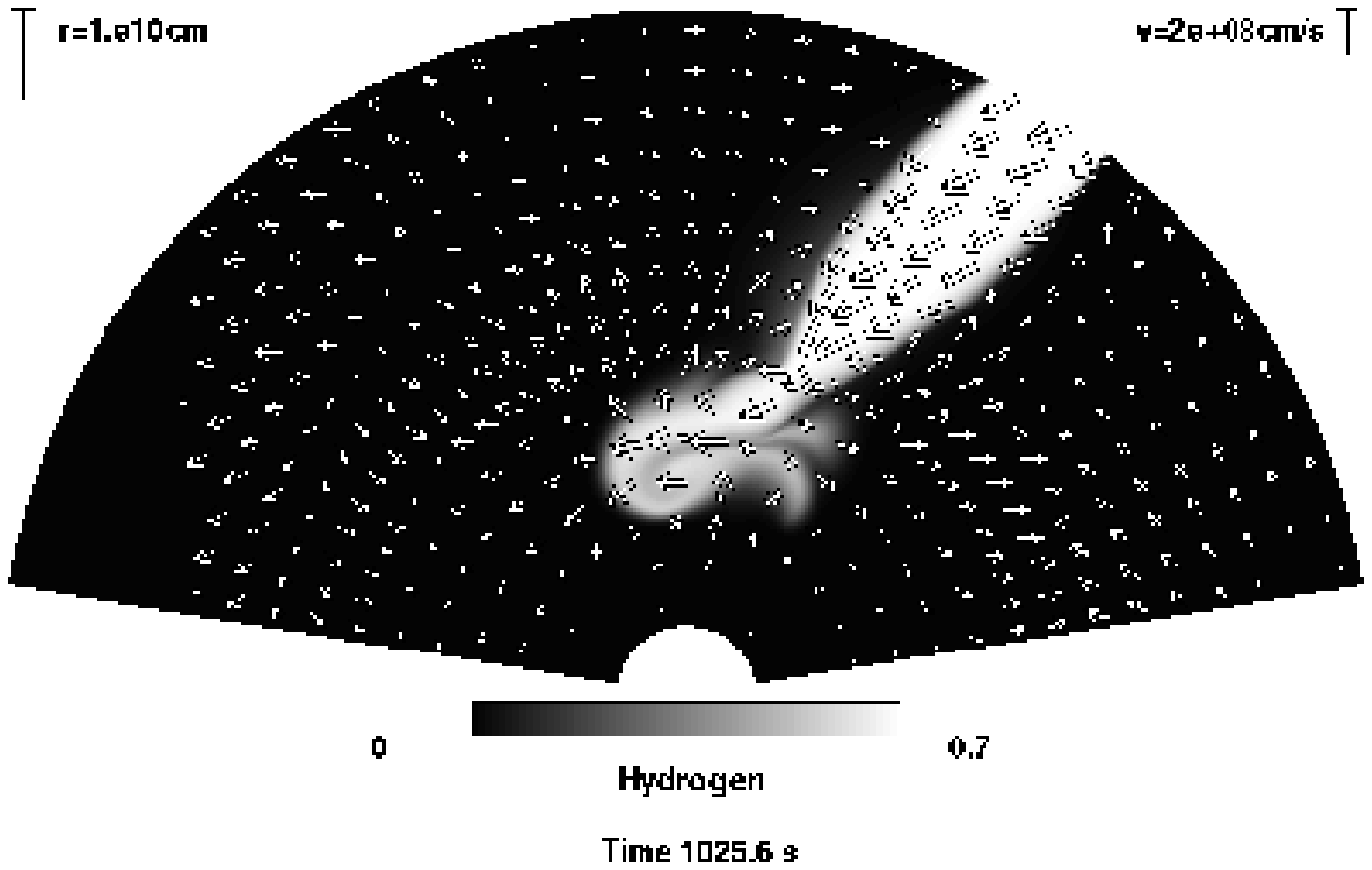}\hspace{2cm}
\includegraphics[scale=0.45]
 	{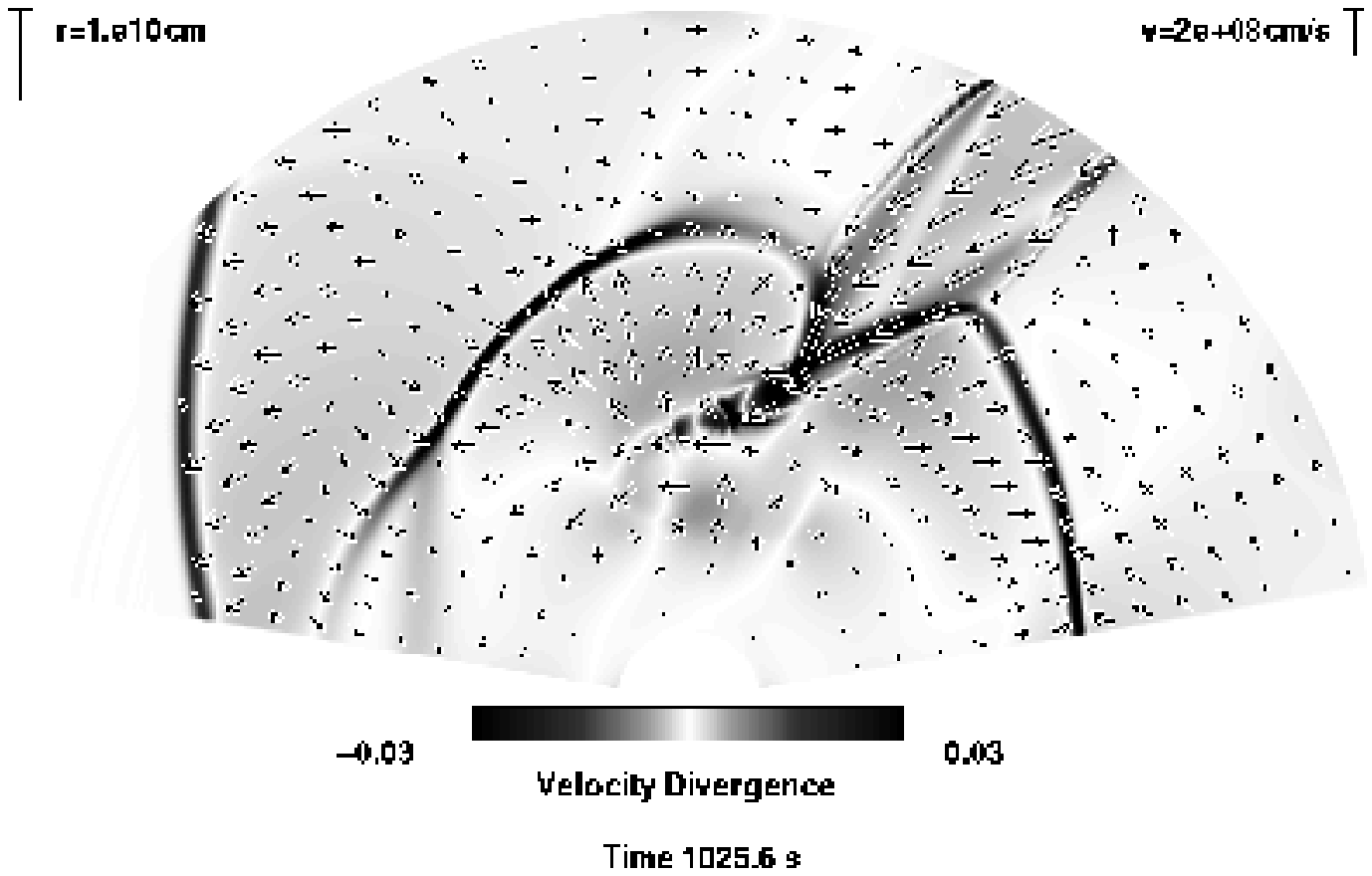} \\
\includegraphics[scale=0.45]
 	{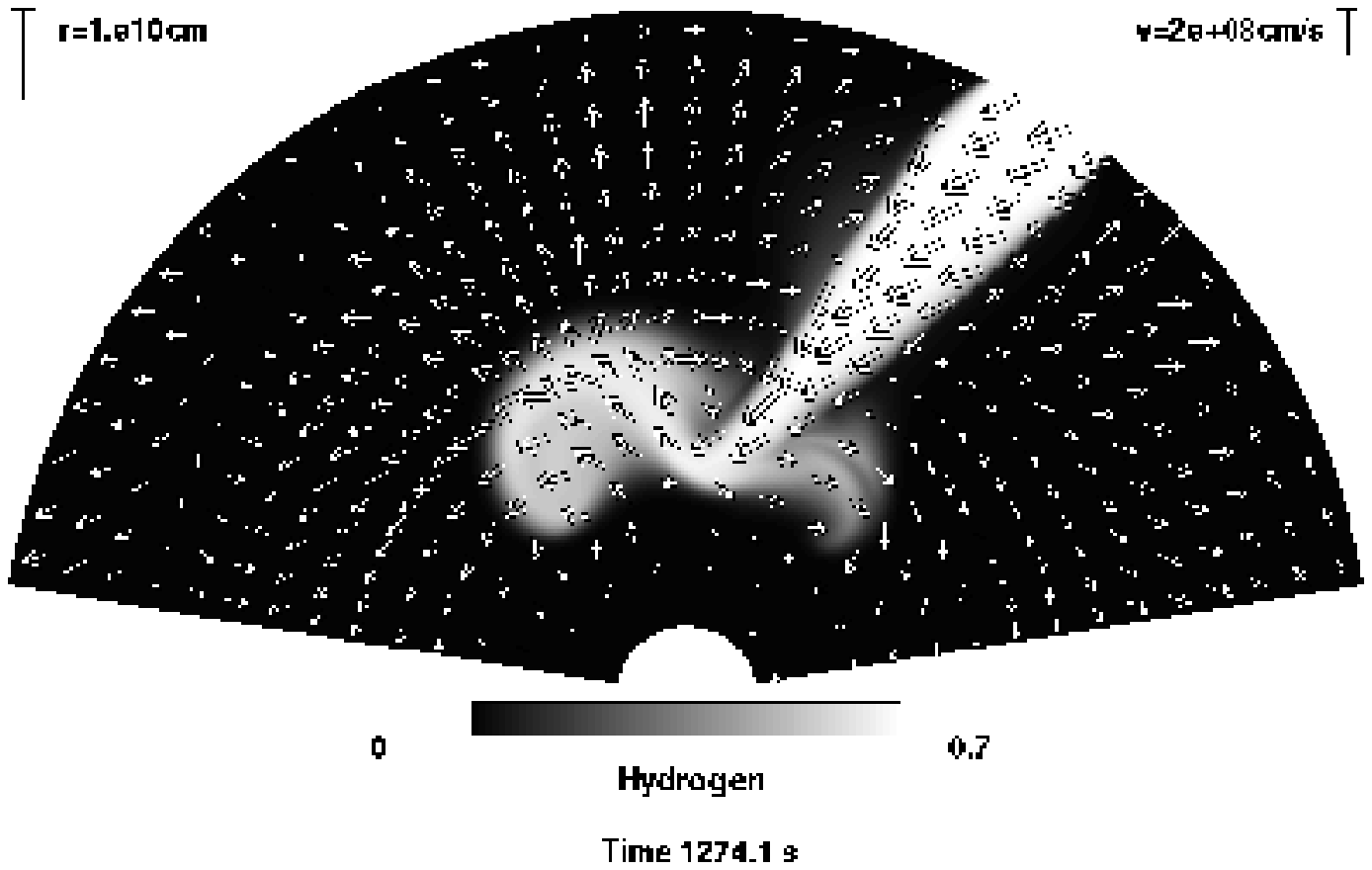}\hspace{2cm}
\includegraphics[scale=0.45]
 	{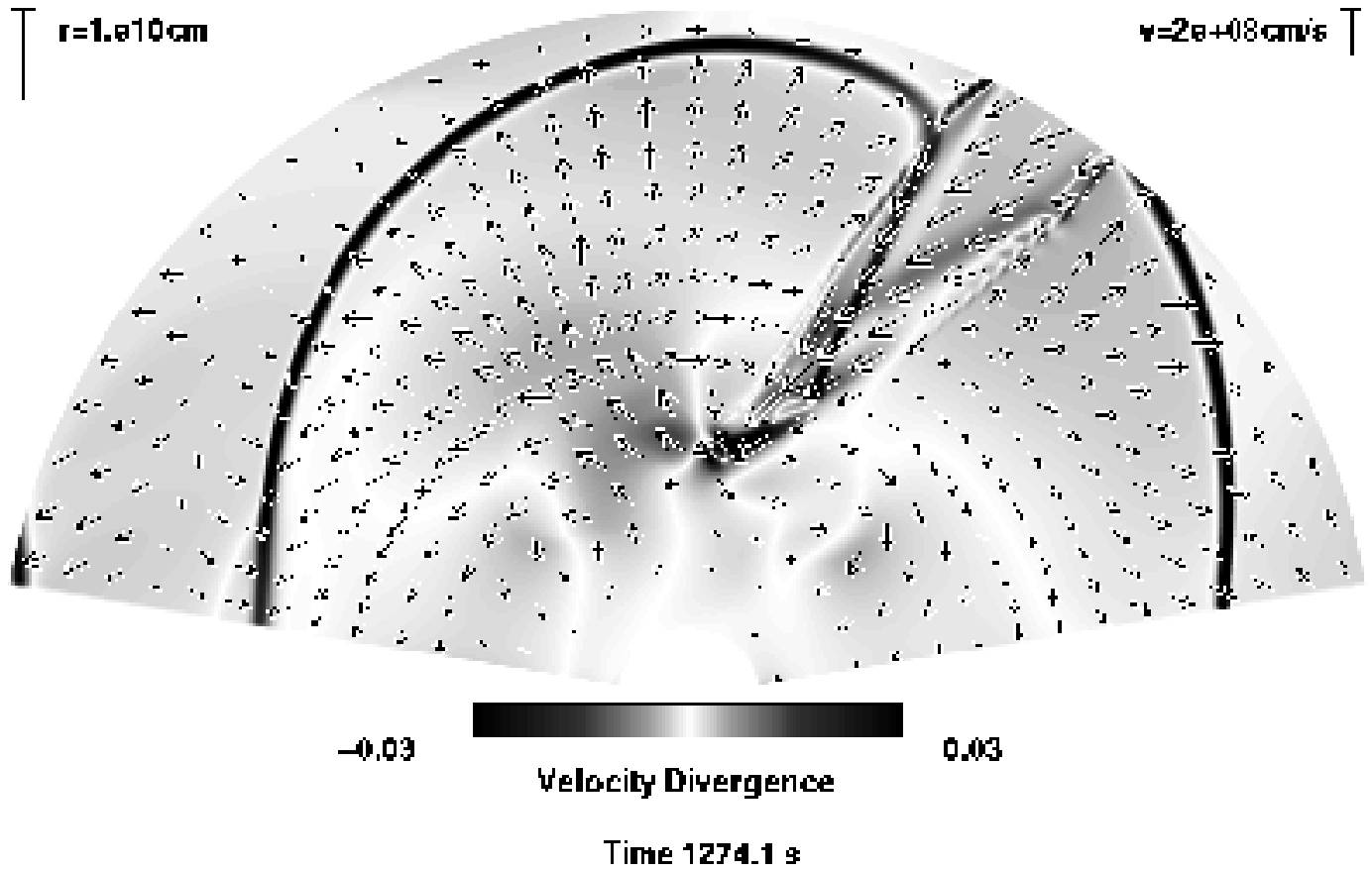} \\
\includegraphics[scale=0.45]
 	{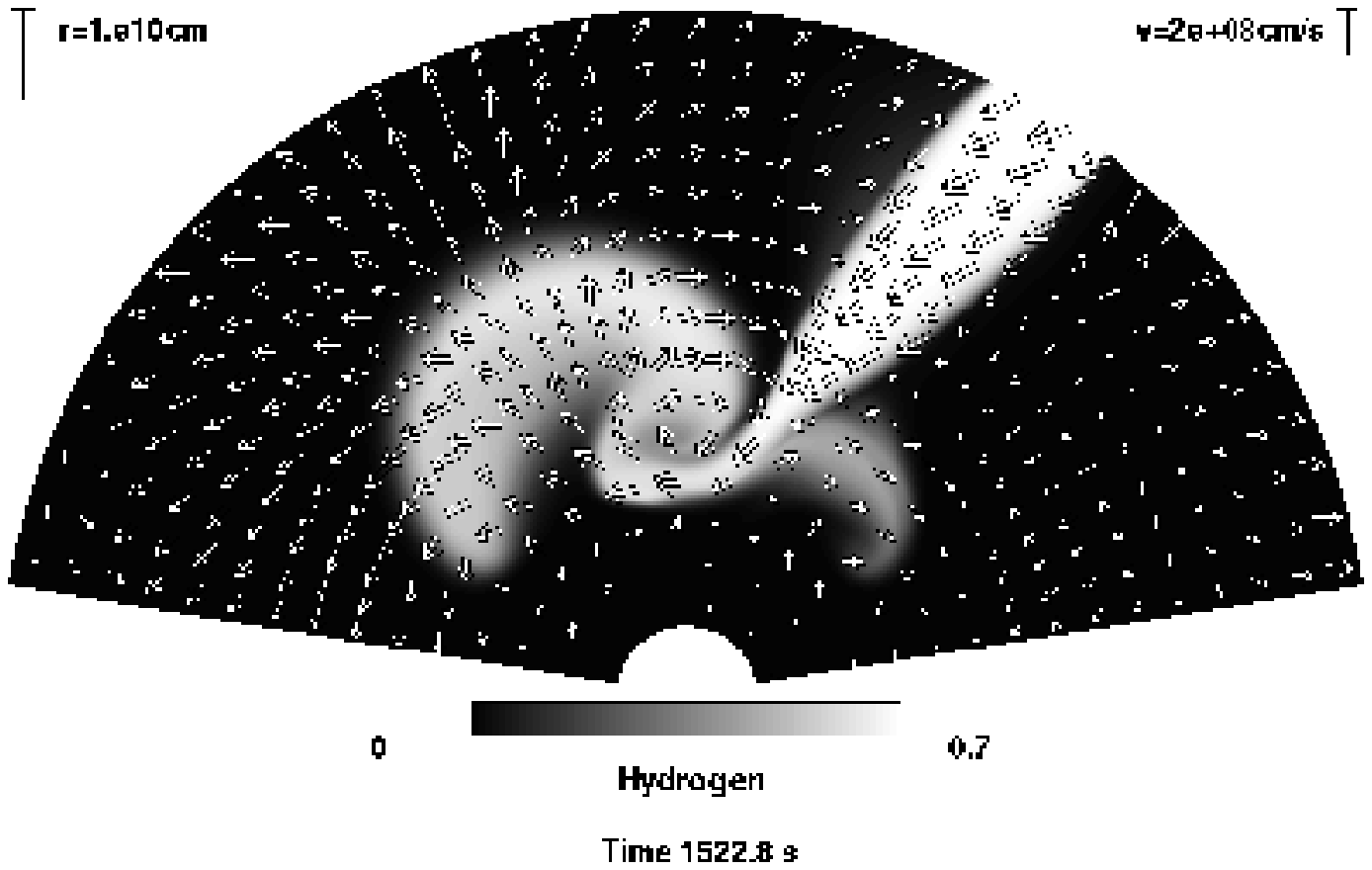}\hspace{2cm}
\includegraphics[scale=0.45]
 	{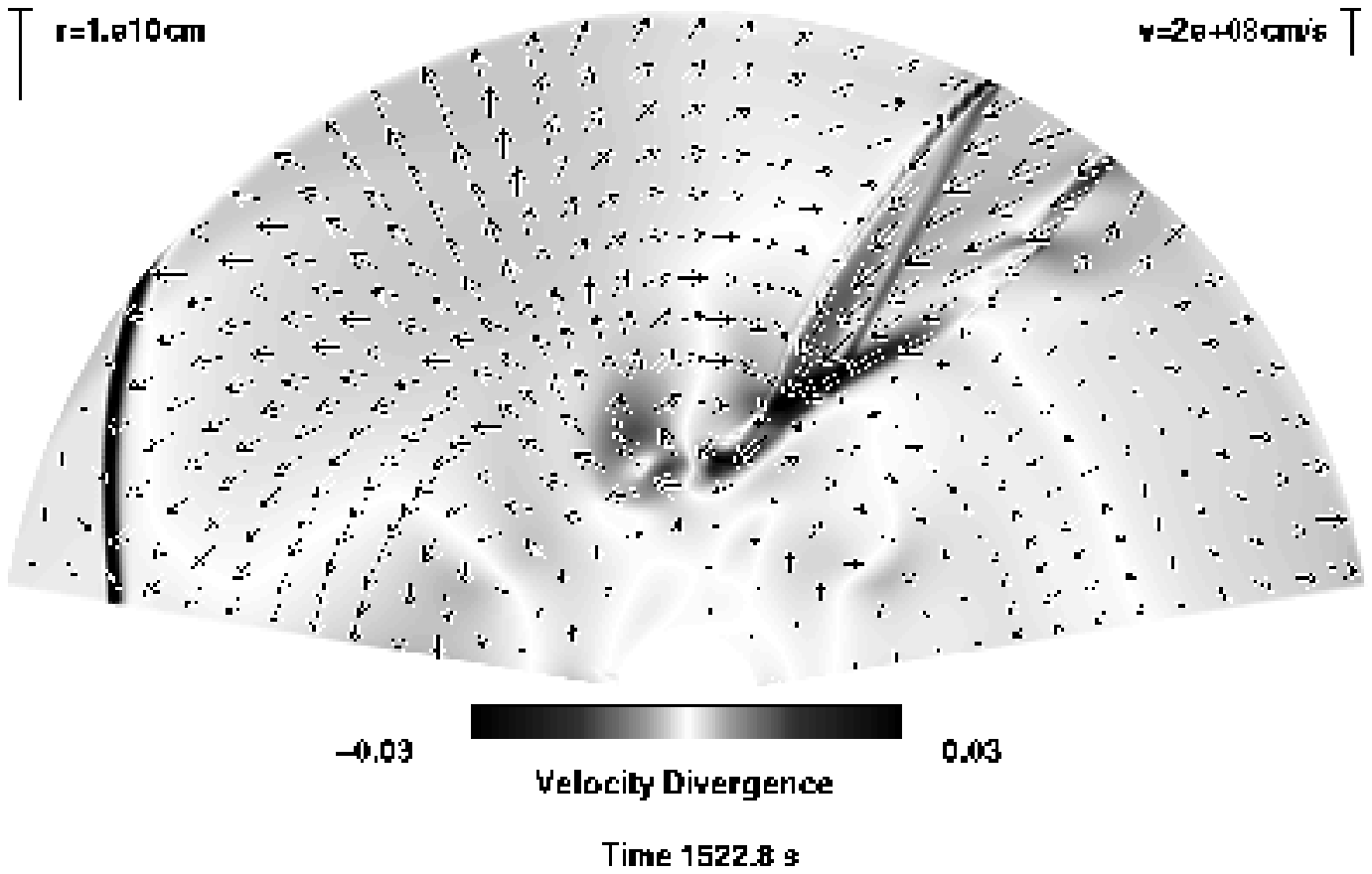}
\caption{\label{fig:initial} The initial stream-core interaction and
the establishment of a steady stream in a frame rotating with the core
(with an angular velocity $\Omega_{\rm frame}=4\times 10^{-4}$\rads).
The left panels show the distribution of hydrogen, the right panels
the distribution of $\bmath\nabla\cdot \bmath v$.  The initial Mach numbers in
the stream, $M_{\rm int}$, and ambient medium, $M_{\rm ext}$ are 8.5 and
3.2, respectively, the density ratio between the stream and the ambient
medium, $\eta_\rho=8$, the outer radius $R_{\rm T} =7.5\times
10^{10}$\,cm, the angular radius of the stream $R_{\rm s} = 0.03 \pi$ and
the initial stream inclination with respect to the radial direction
$\Theta=13.5\degr$.  The velocity field is scaled according to
$\log(v/10^6\cms)$.}
\end{figure*} 

Figure~\ref{fig:initial} shows the initial interaction of a stream with
a massive core and illustrates the development of a stationary stream.
The parameters for this simulation are characteristic for a 1\Msun\ star
filling its Roche lobe inside the envelope of a very evolved 20\Msun\ 
red supergiant (after helium core burning). The panels on the left show
the hydrogen abundance, while the panels on the right show the divergence 
of the velocity field. The latter was chosen because it shows shock 
structures particularly clearly.

The top panels show the stream just before its `impact' with the core
(569\,s after the beginning of the simulation), while the second set
of panels is close to the point of impact (at 778\,s; note the bow
shock in the right panels). Immediately after the impact, stream
matter bounces off the core, where most of it is just reflected by the
core and continues to move in the forward direction (i.e.
counter-clockwise), 
but some of it is pushed backwards, attaining an
angular velocity component opposite to the rotation of the core.  As
material that has been stopped by the core is pushed by material in
the stream following from behind, it starts to flow up again, driving
two powerful shocks on the front and back side of the stream into the
envelopes. These shocks compress the stream significantly.  Once they
have left the domain of the calculation, the stream has attained 
essentially a stationary configuration, where the point of deepest penetration
($R\simeq 2\times10^{10}\,$cm) and the stream shape no longer change
significantly. The overall flow pattern also becomes more-or-less
stationary, where all the matter leaving the stream in the core impact 
region flows up again, being vigorously mixed with helium from the core in
the process.

\subsection{Mechanisms for stream dissipation}

The  main objective of our calculation is to determine how deep
the stream can penetrate into the core of a massive star rather than
the initial transient behaviour. In the quasi-stationary situation
this depends mainly on the entropy that is generated in the stream-core
interaction.

\subsubsection{Pressure discontinuity and entropy generation}

\begin{figure*}
\includegraphics[scale=0.5]
 	{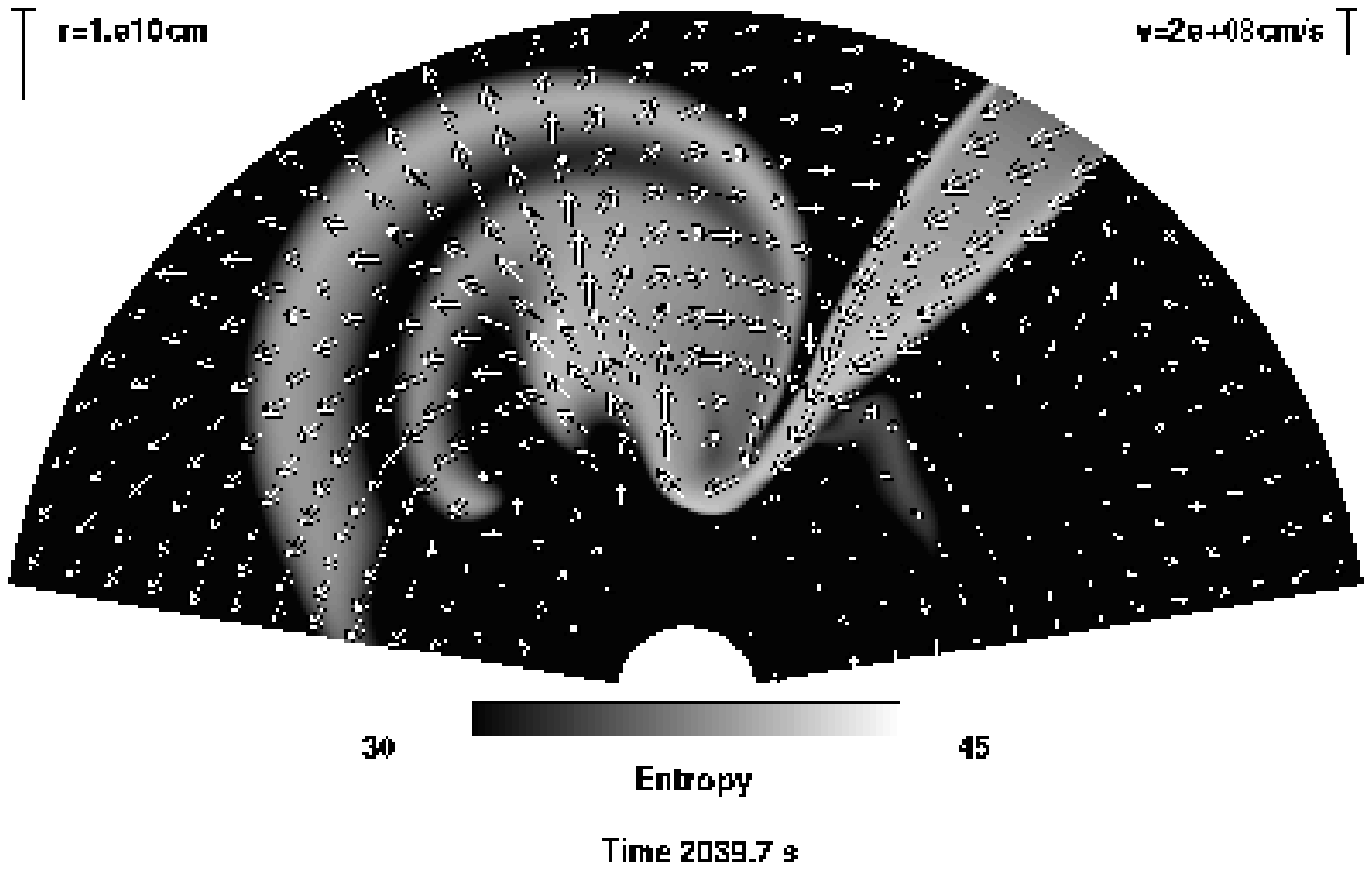}\hspace{1cm}
\includegraphics[scale=0.5]
 	{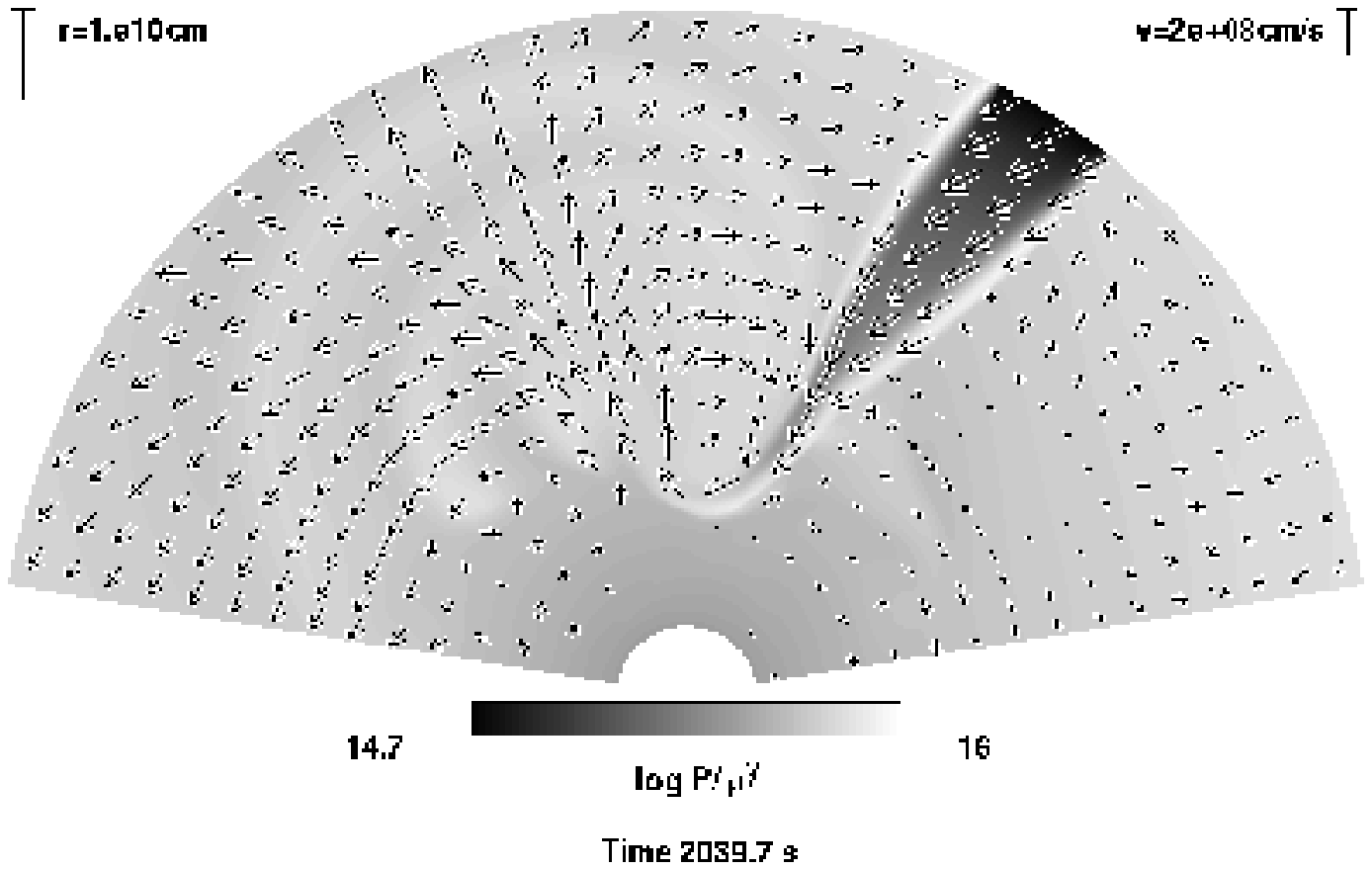} \\
\includegraphics[scale=0.5] 
 	{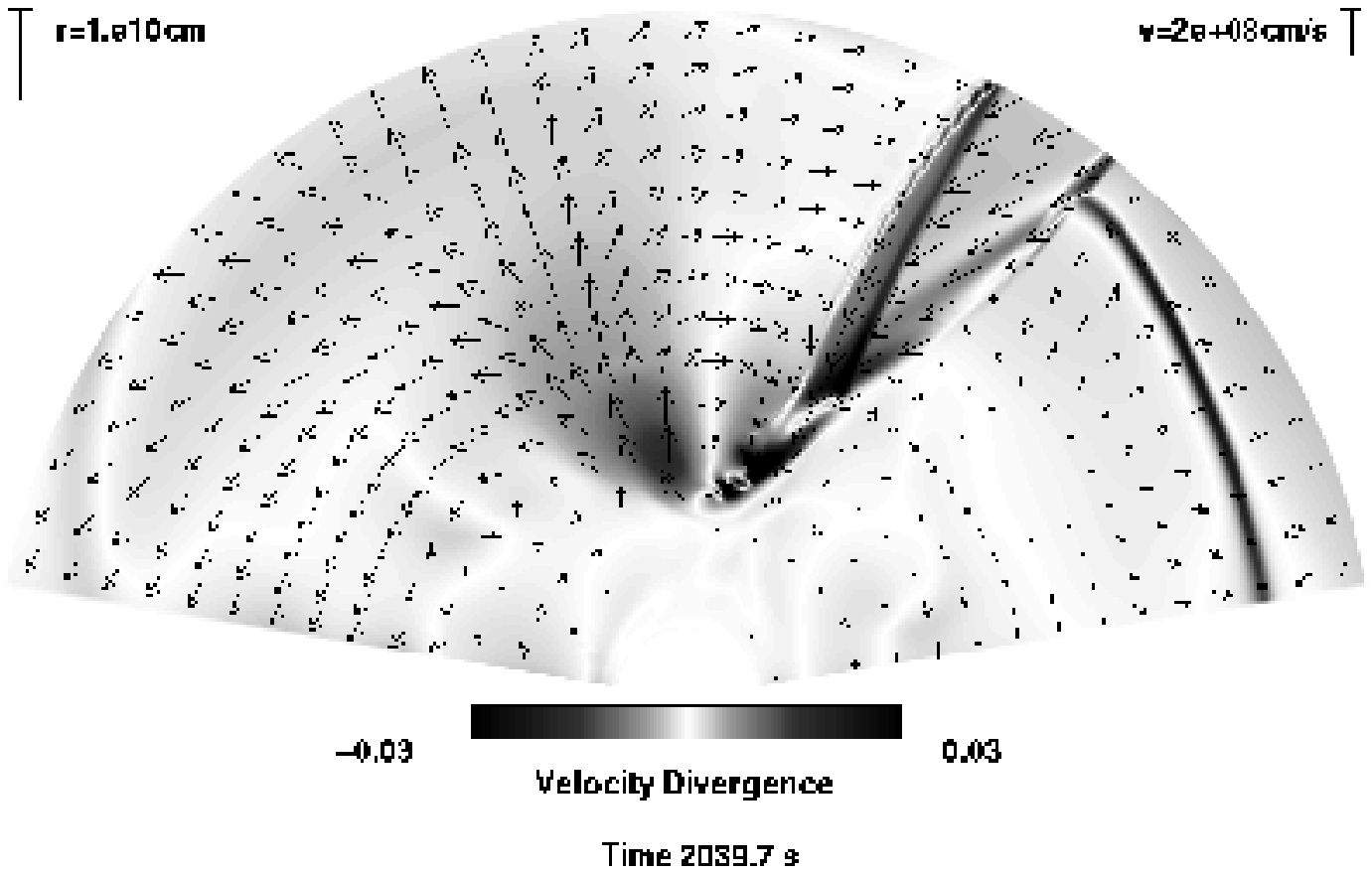}\hspace{1cm}
\includegraphics[scale=0.5]
 	{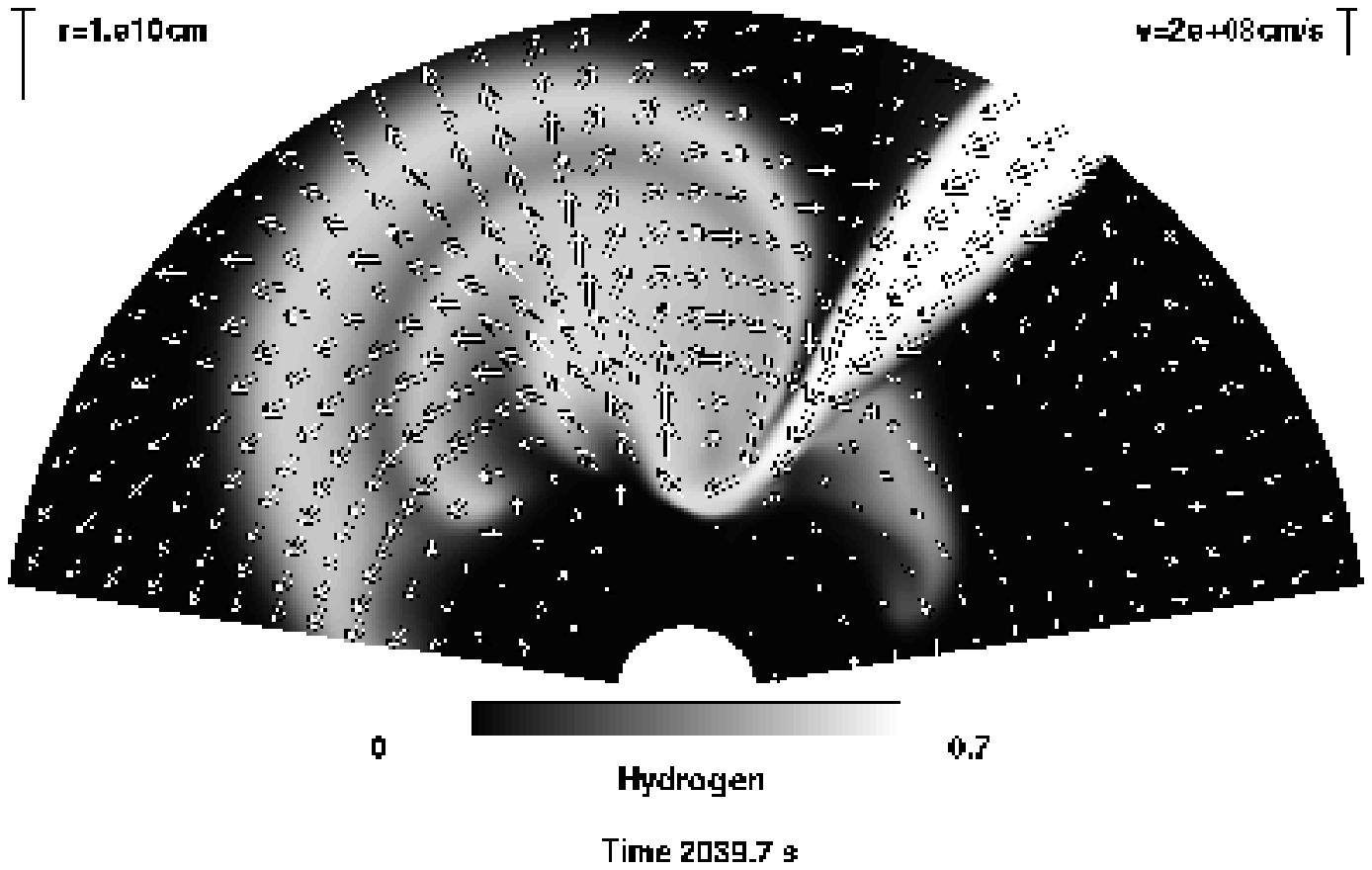} \\ 
\includegraphics[scale=0.5]  
 	{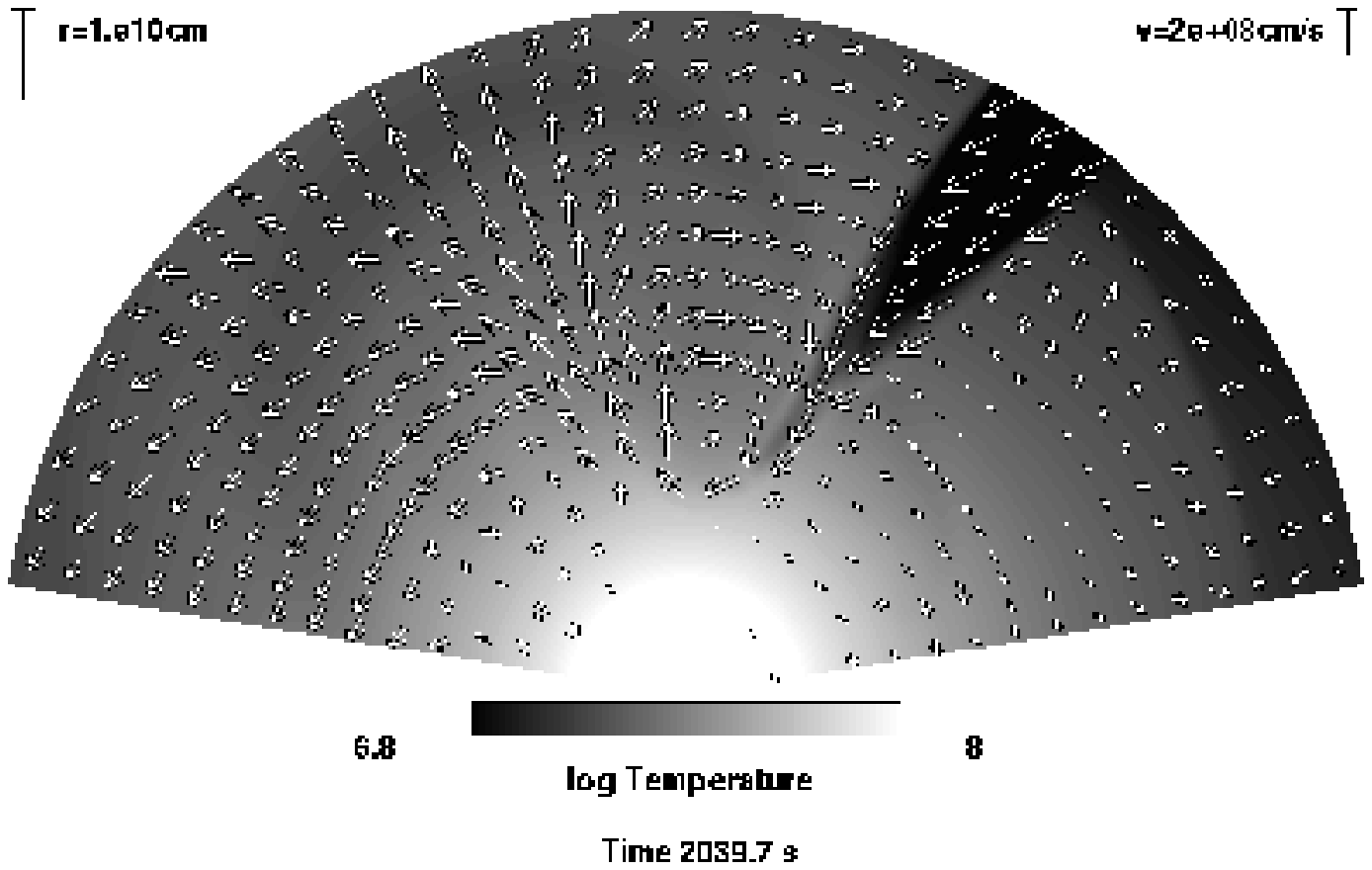}\hspace{1cm}
\includegraphics[scale=0.5]
 	{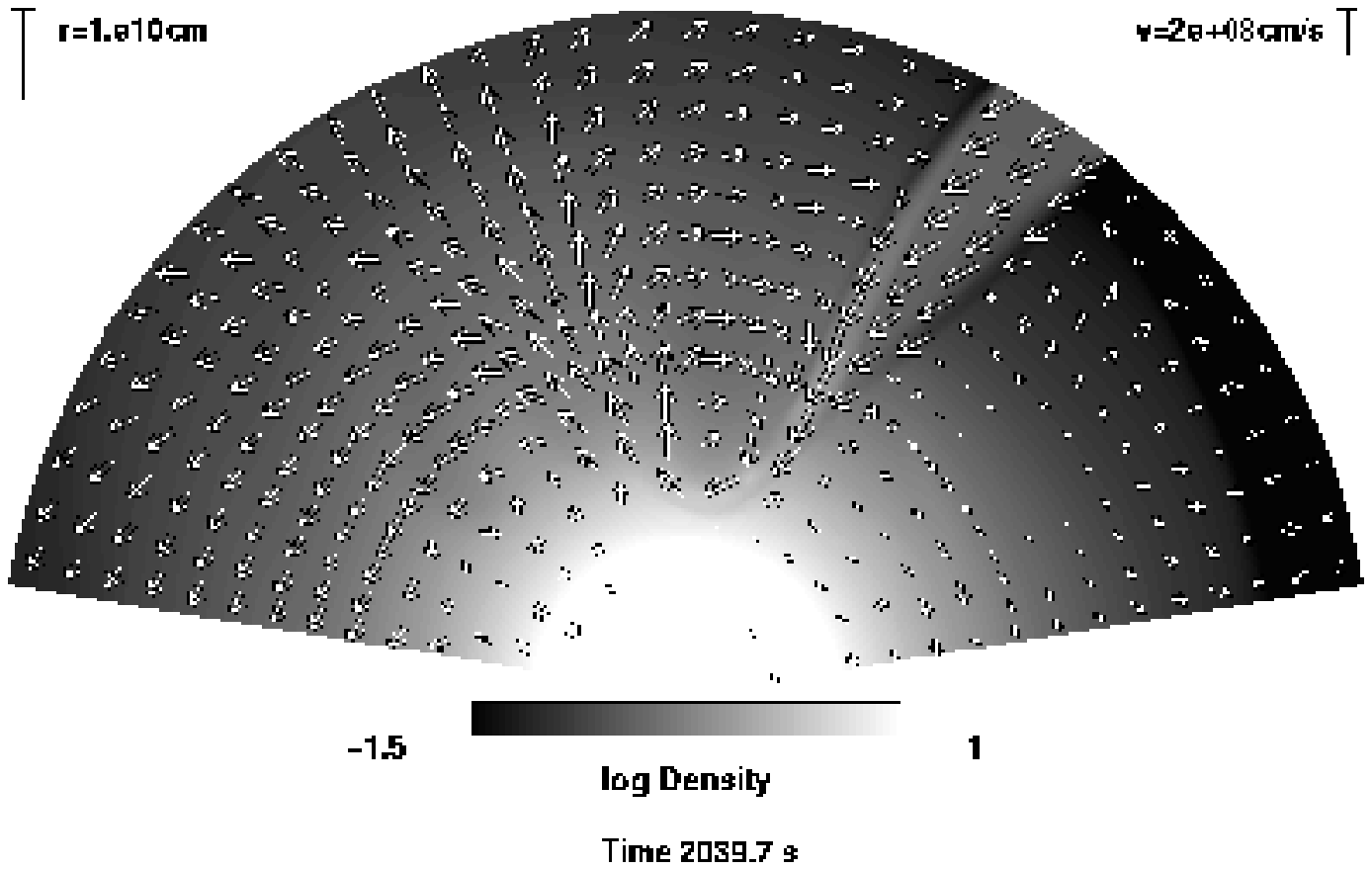} 
\caption{ \label{fig:pr_disc} 
Characteristic properties (entropy, $\log P/\rho^{\gamma}$ ,
divergence of velocity, hydrogen mass fraction, density and temperature)
for a stationary inclined stream (with the same parameters as
in Fig.~\ref{fig:initial}). The radius of deepest penetration 
is $2.0\times 10^{10}\,$cm in the simulation. At that depth the
temperature in the stream is $6\times 10^7\,$K and the 
density $3\gcm$.
}
\end{figure*}

While the stream is in lateral pressure balance with the ambient 
matter at an early stage of infall, it becomes increasingly unbalanced
as the velocity increases and becomes significantly supersonic relative
to the ambient medium. 
The interaction between the stream and the ambient matter
due to the jump in pressure can be treated in a simplified way as a Riemann
problem (see e.g. LeVeque et al.\ 1998),
a solution of which is a combination of shock
and rarefaction waves. In the case where the stream expands,
the rarefaction wave, propagating into the stream material, does not
change the stream entropy. However, compression of the stream by
the ambient matter, in the form of shocks moving into the stream,
generates entropy inside the shocks.
 
For a strong shock, where the  pressure jump $P_{\rm amb }/P_{\rm s } 
\gg 4$, the coefficient of entropy change, i.e. the ratio of the entropy
of the shocked stream material, $S_{\rm ss}$, to the initial entropy 
in the stream, $S_{\rm s}$, can be estimated as

\beq
K_{\rm S } = S_{\rm ss }/ S_{\rm s } \approx \const \cdot P_{\rm amb }/P_{\rm s } 
\eeq
\noi \cite{Zeld}.

In our case, we rather expect the development of weak shocks.
Then an estimate for $K_{\rm S } $ can be written as 

\beq
K_{\rm S } \approx 1 + \const \cdot \eta_{\rho}^{\gamma-1} 
\left ({\frac {M_{\rm int }} {M_{\rm ext }}} \right ) ^2 
(P_{\rm amb }/P_{\rm s } -1)^3 \ 
\label{eq:ks}
\eeq

\noi \cite{Zeld}.
The pressure jump  $P_{\rm amb }/P_{\rm s }$ changes 
with the distance to the primary core. In the case of a power-law
parametrisation of the ambient pressure, equation~(\ref{eq:ks}) 
predicts that models
with smaller power-law indices will result in smaller pressure jumps
for the same initial stream properties and hence smaller values
for $K_{\rm S}$ and that this coefficient increases as the power-law
index increases.
  
In Figure~\ref{fig:pr_disc} we present some of the key characteristics
of a stream with the same parameters as in Figure~\ref{fig:initial},
once a more-or-less stationary flow pattern has been established.  Note in
the top two panels that the entropy in the core of the stream is
initially constant and that all the entropy generation takes places in
two strong shocks at the stream boundaries (clearly seen in the
$\bmath \nabla\cdot \bmath v$ panel). All of the stream matter has to pass
through the intersection of these shock structures, where it
essentially attains its final entropy and becomes comparable to the
entropy of the ambient material, at least in the case of the
 $\log P/\rho^{\gamma}$. For the entropy the difference between ambient
and stream matter is largely caused by the difference in molecular
weights and its subsequent change is due to mixing of these two media
rather than dissipation. At the point of deepest penetration ($\sim 2\times
10^{10}\,$cm), the stream matter has a temperature of $6 \times 10^7\,$K
and a density of $\sim 3 \gcm$, high enough for nuclear
burning (see \S~5.4).

\subsubsection{Kelvin-Helmholtz instabilities and the stream width} 

\begin{figure}
\centering
\includegraphics[scale=0.5] {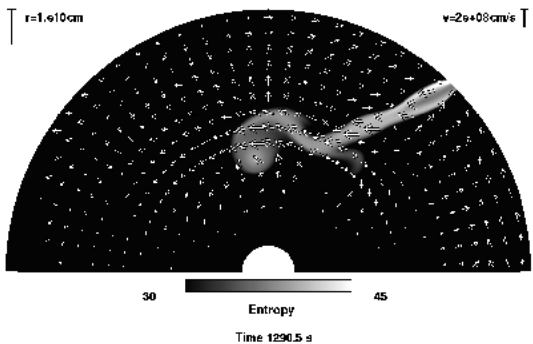}
\caption{\label{fig:serp} Example of a Kelvin-Helmholtz instability
experienced by a narrow stream with $ R_{\rm s } = 0.005 \pi$.
($M_{\rm int } = 8$, $M_{\rm ext} = 3.25$, $\eta_\rho = 8$, $R_{\rm T } 
= 7.5\times  10^{10}{\rm cm }$, $\Omega_{\rm frame }=4\times 10^{-4}\rads$, 
$\Theta=18\degr$.) The stream penetrates to a depth of 
$2.7\times 10^{10}\,$cm.}
\end{figure}

The K-H instability  affects both the stream shape
and the stream width. For small relative velocities, the stream is
``sausage''-shaped and becomes  increasingly ``snake''-shaped
as the relative velocity increases (see Fig.~\ref{fig:serp}).
It becomes more important as the stream narrows,
since perturbations of a given wavelength are more important for a
narrower flow. It therefore
determines the minimum stream width and thus limits the depth of penetration.
In particular, a stream that is very narrow initially will penetrate less deep
than a stream with a similar parameters like velocity, density and pressure,
but with a larger mass-flow rate and hence larger initial stream width.

\subsection{The effects of rotation}

\subsubsection{A rotating medium}

\begin{figure}
\includegraphics[scale=0.5]
 	{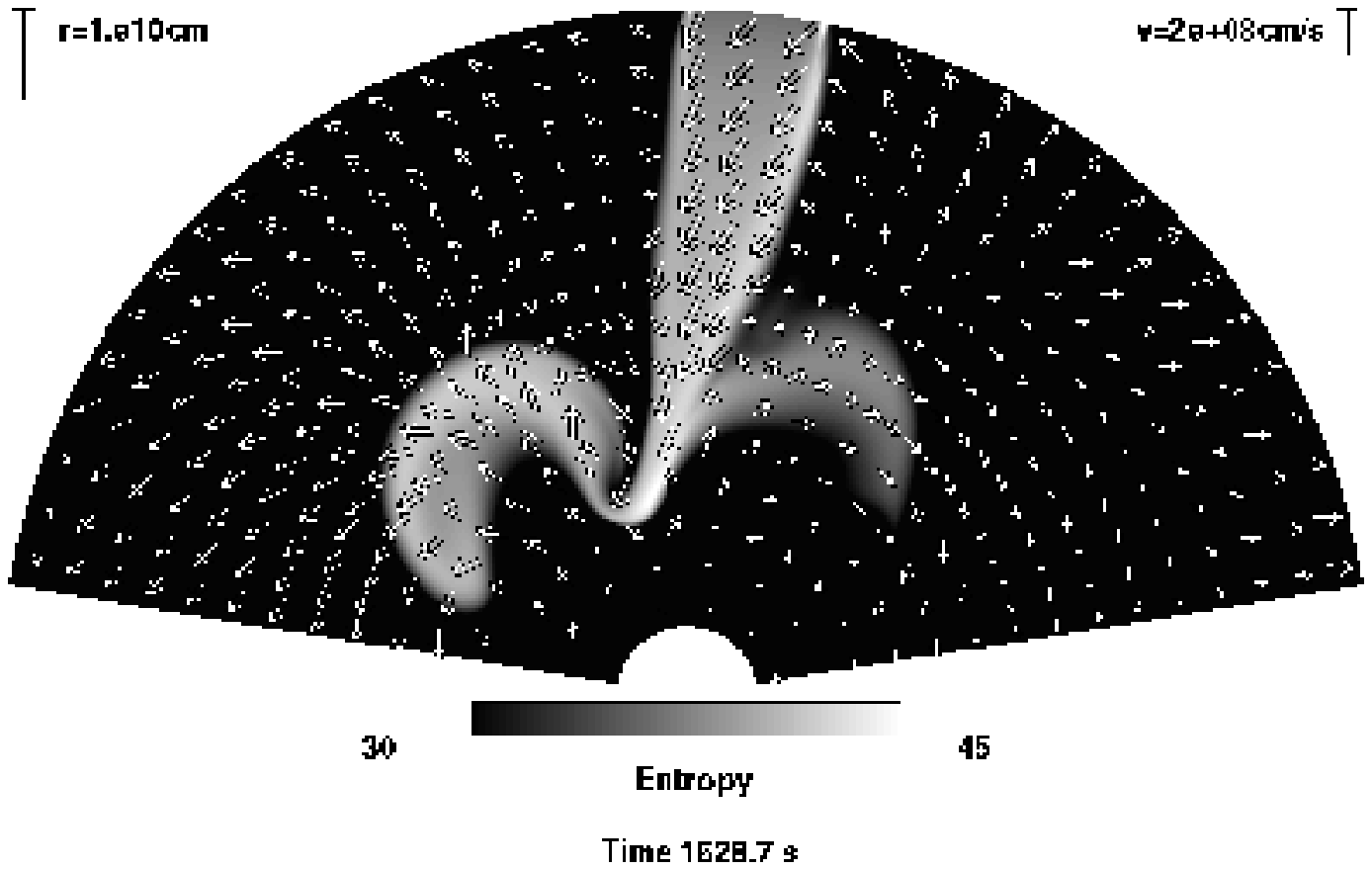}
\includegraphics[scale=0.5]
 	{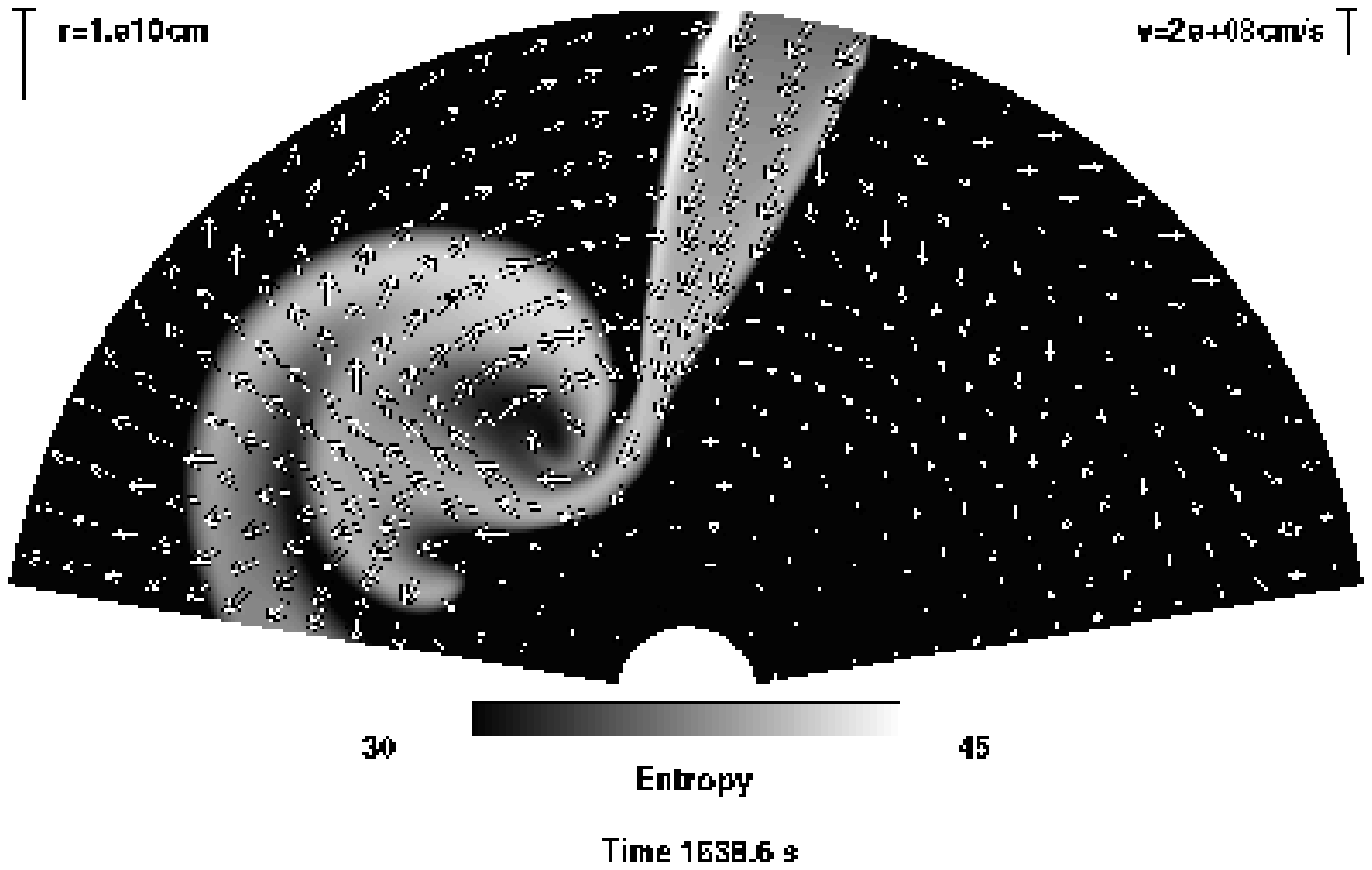}
\caption{ \label{fig:rot} 
Examples for streams penetrating into a medium rotating relative
to the binary. The stream rotates in an anti-clockwise and a clockwise
direction relative to the medium in the top and bottom panels, 
respectively, with a relative velocity  $\Omega_{\rm shift } = 10^{-4}\rads$.
The initial positions of the stream centre in the azimuthal direction 
were taken to be $0.3\pi$ and $0.6\pi$ in the
two cases, respectively. The other stream parameters are the same as
in Figure~\ref{fig:pr_disc}
($M_{\rm int} = 8.5$, $M_{\rm ext} = 3.2$, $ \eta_\rho = 8$, $ R_{\rm T } 
= 7.5\times  10^{10}\,$cm, $R_{\rm s } = 0.03 \pi$,
$\Omega_{\rm amb }=4\times 10^{-4}\rads$, 
$\Theta=13.5\degr$). The radii at the point of deepest 
penetration are 1.75 and $2.25\times 10^{10}\,$cm, 
respectively.
}
\end{figure}

In some cases, in particular near the end of the merging process, the
envelope surrounding the spiraling-in components may no longer remain
in co-rotation with the binary (either because the spiral-in time-scale
becomes too short or the region surrounding the primary core expands
rapidly). Then the stream feels an additional force due to the moving
external medium.  This affects the stream's trajectory, in particular
the angle of incidence, and thereby the penetration depth. This is
illustrated in Figure~\ref{fig:rot} which shows two stream
calculations with the same parameters as in Figure~\ref{fig:pr_disc}
but where the medium is assumed to rotate with an angular velocity
$\Omega_{\rm shift} = 10^{-4}\rads$ relative to the frame of the binary
(rotating with angular velocity $\Omega$) either in the forward or in
the backward direction.  The frame of the medium (in which the
calculation is performed) then rotates with a velocity $\Omega_{\rm
amb} = \Omega \pm \Omega_{\rm shift}$. In this frame, the stream
itself rotates either anti-clockwise (top panel) or clockwise (bottom
panel). As the stream is being pushed backwards by ambient matter
moving against its direction of rotation, its trajectory steepens and
consequently the penetration depth increases (top panel), while the
opposite happens when the stream is pushed from behind (bottom panel).
As a consequence the stream penetrates deeper to a radius of $\sim
1.75\times 10^{10}\,$cm instead of $\sim 2\times 10^{10}\,$cm in the
standard case (Fig.~\ref{fig:pr_disc}), while it only reaches a depth
of $\sim 2.25\times 10^{10}\,$cm in the opposite case. The
corresponding change in the stream temperature at the deepest point is
only a few per cent. Note also that for the stream rotating
anti-clockwise, the entropy generation at the back side of the stream,
where the dynamical pressure is smaller than at the front side (and hence
has a larger jump in pressure), is larger than for stationary inclined stream.

\subsubsection{Core rotation}

\begin{figure*}
\includegraphics[scale=0.5]
 	{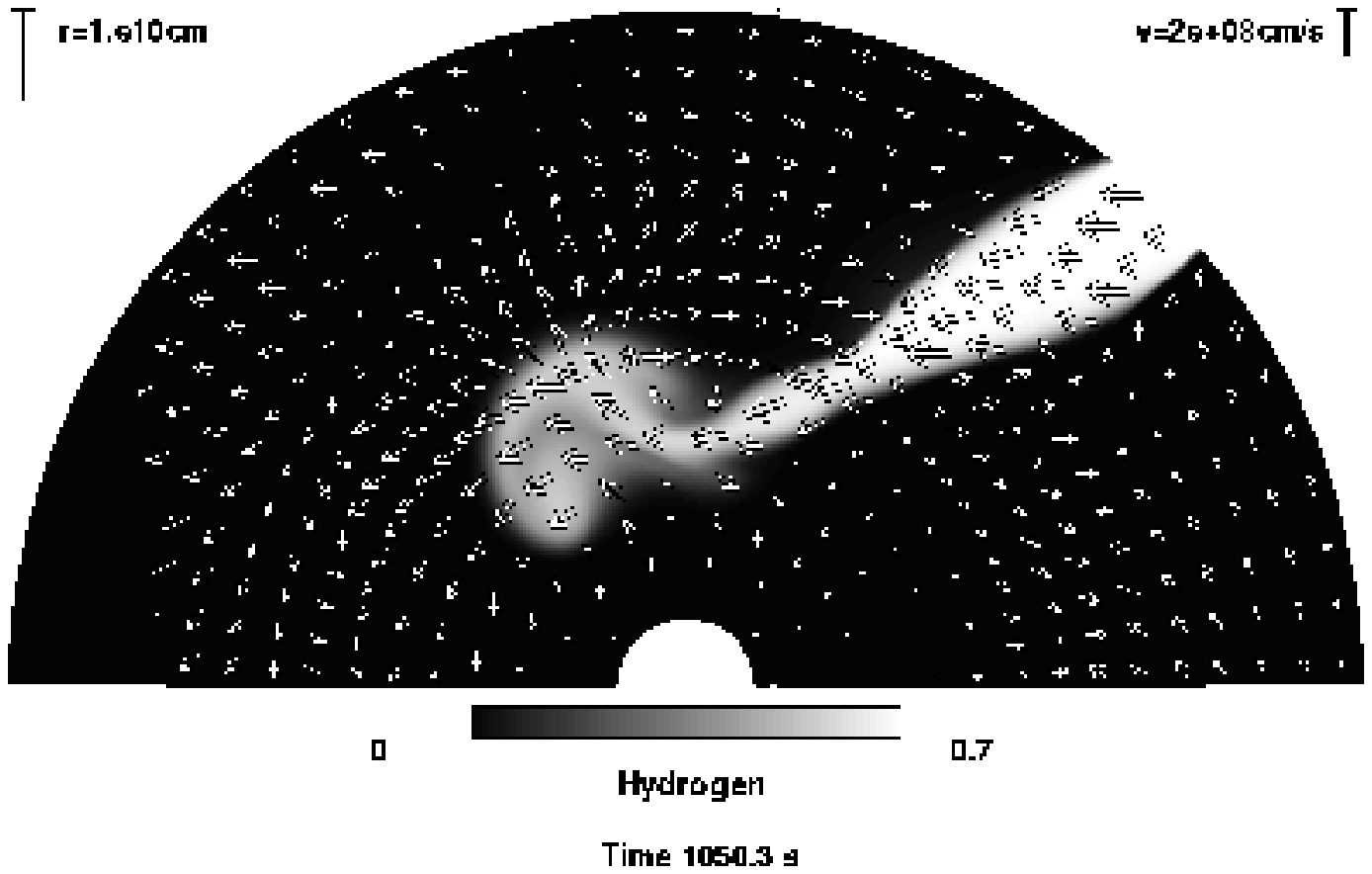}\hspace{1cm}
\includegraphics[scale=0.5]
 	{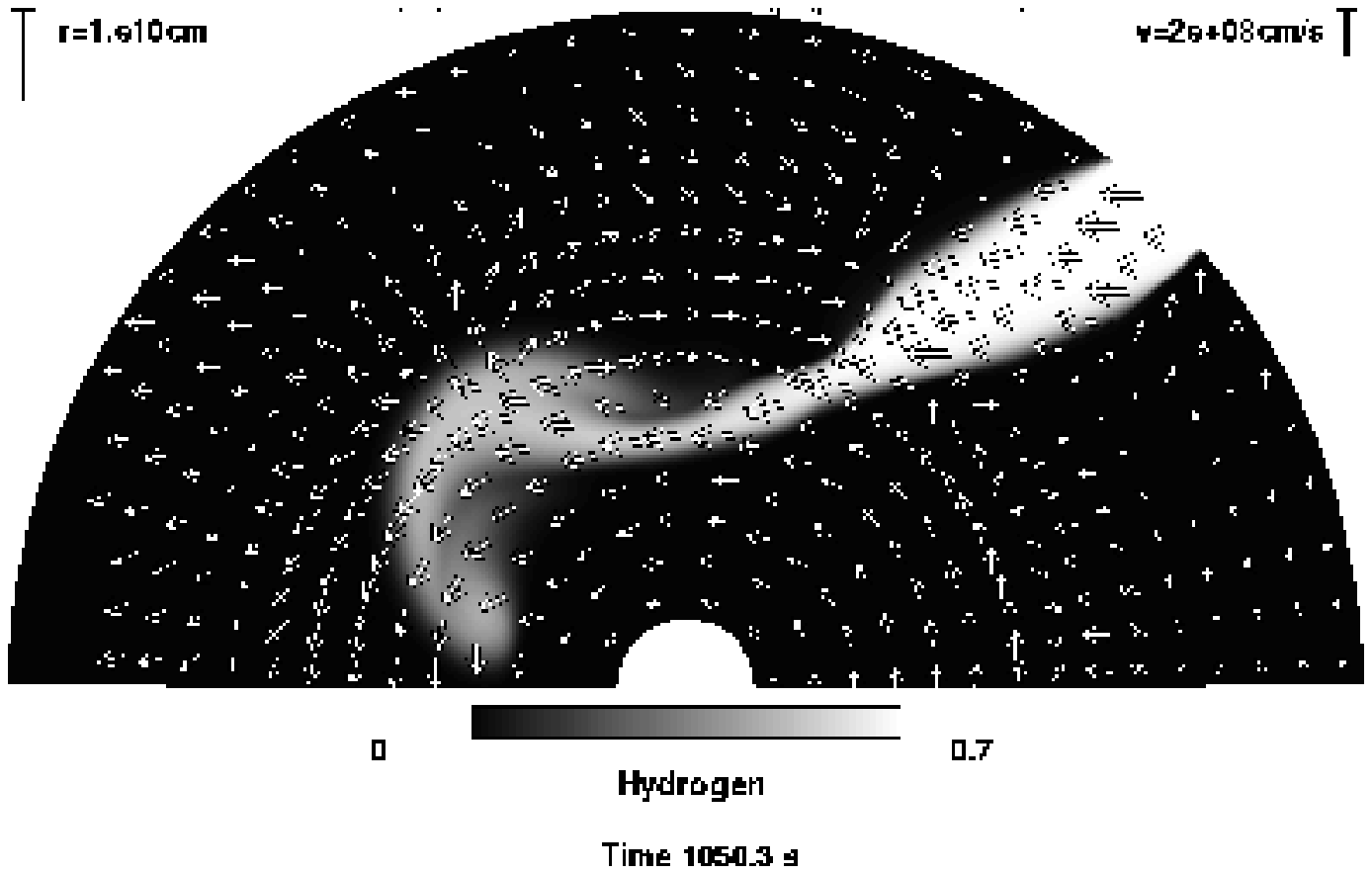} \\
\includegraphics[scale=0.5] 
        {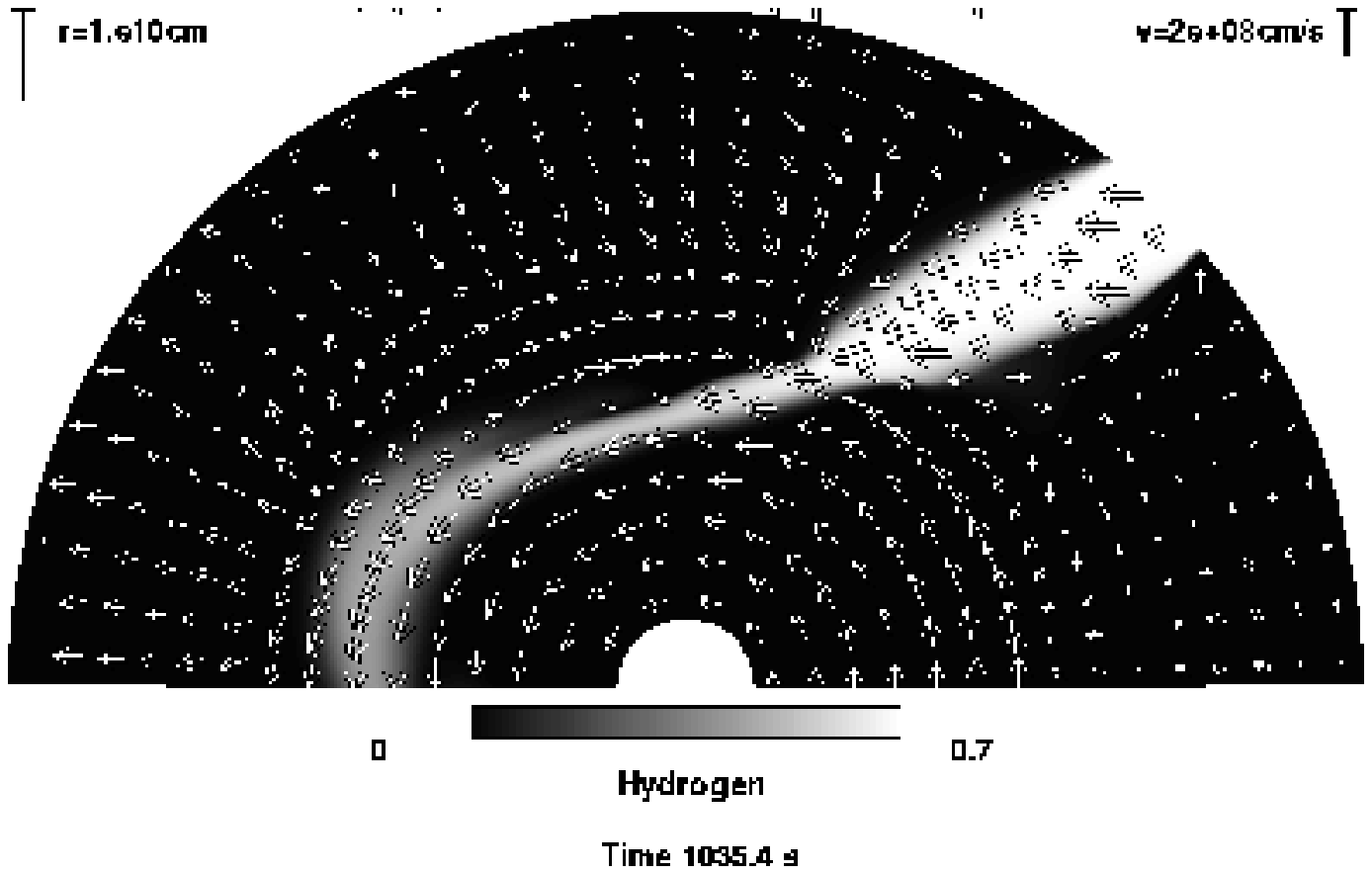}\hspace{1cm}
\includegraphics[scale=0.5]
 	{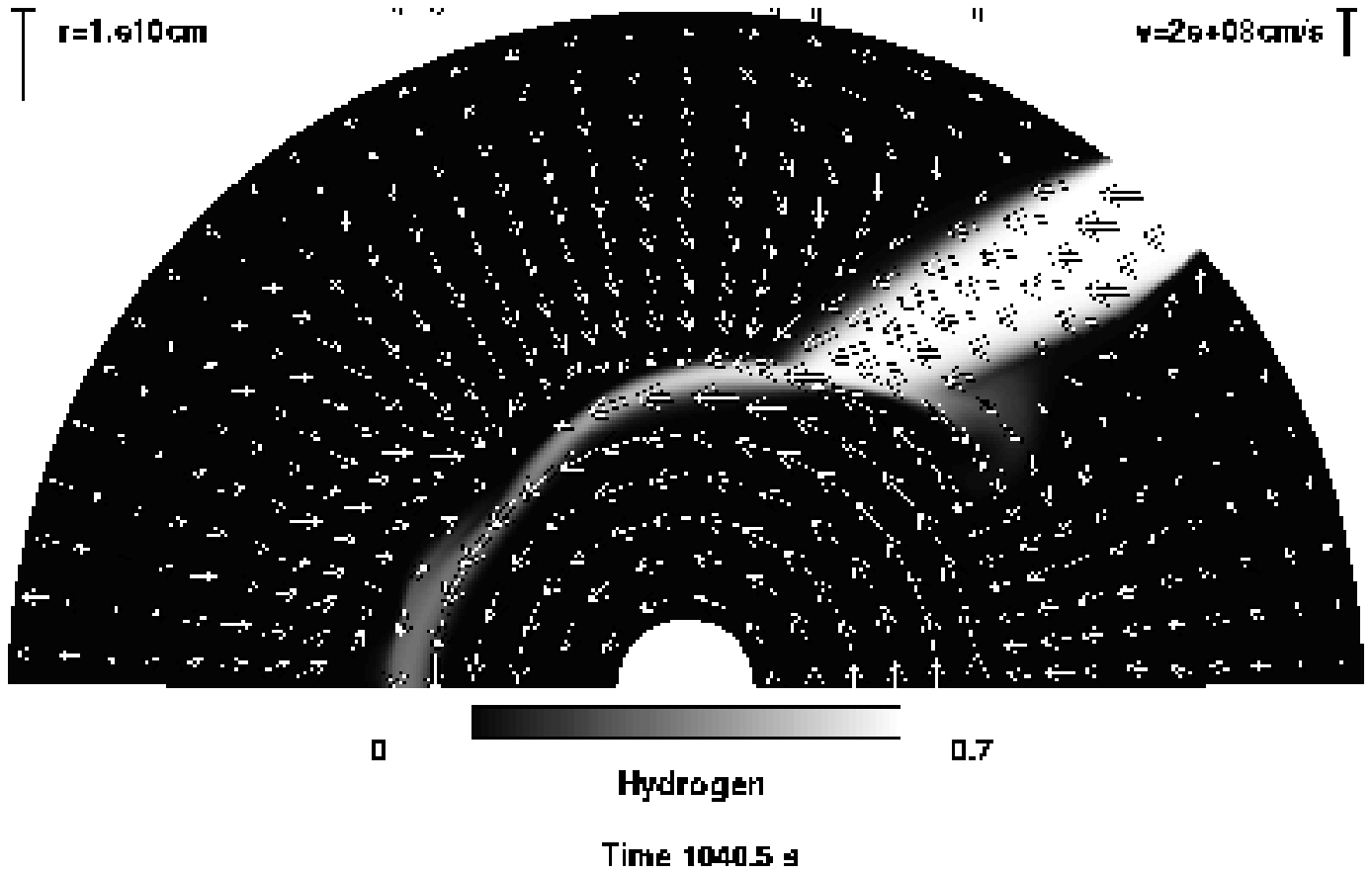}
\caption{ \label{fig:core} 
Stream simulations illustrating the effects of rotation of the primary
core (in a frame rotating with the binary). 
In the four calculations, the core is assumed to rotate with constant
angular velocities of $\Omega_{\rm core } = 0$ (top left), 0.001 (top right),
0.0025 (bottom left) and $0.005\rads$ (bottom right), respectively, up
to a radius $4\times 10^{10}\,$cm. The other stream parameters are
$M_{\rm int} = 9.8$, $M_{\rm ext} = 3.7$, $\eta_\rho = 8.4$, 
$R_{\rm T } = 7.5\time 10^{10}\,$cm, $R_{\rm s } = 0.025 \pi$,
$\Theta=18\degr$.
}
\end{figure*}

\begin{figure}
\includegraphics[scale=0.47]{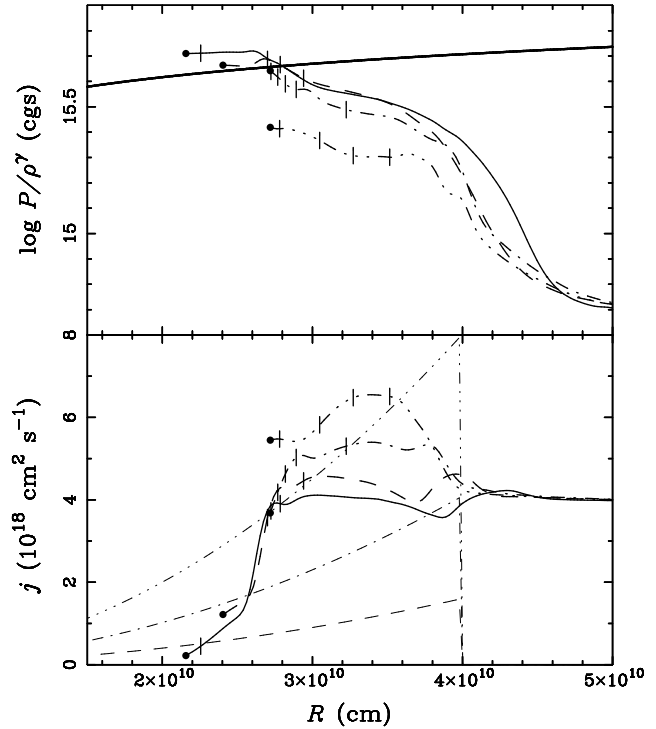}
\caption{ \label{fig:core2} 
The logarithm of the
adiabatic constant ($s=P/\rho^{\gamma}$, top panel) and 
specific angular momentum (bottom
panel) near the centre of the streams as a function
of penetration depth for the simulations in Figure~\ref{fig:core} 
with angular velocities $\Omega= 0$ (solid curves), 0.001 (dashed
curves), 0.0025 (dot-dashed curves) and $0.005\rads$ (dot-dot-dot-dashed
curves). The corresponding angular-velocity profiles for the ambient 
core matter are shown in the bottom panel as lighter curves with
the same line style. The thick solid curve in the top panel
refers of $\log s$ in the ambient matter.
The filled circles at the end of individual curves mark the points of deepest
penetration. The ticks along each curve indicate particular values
of the central stream hydrogen abundance by mass 
(0.6, 0.5, 0.4, 0.3 from right to left).
}
\end{figure}

As the stream interacts with the core, it injects not only matter but
also angular momentum into the core, spinning it up in the process.
Since the core only has to accrete a small fraction of the angular
momentum available from the secondary to be spun up to critical
rotation, we may expect that this itself will affect the dynamics of
the interaction.  To demonstrate this, we performed four simulations
for streams with the same initial parameters and similar core
properties except that we varied the rotation of the core from 0 to
0.005\rads\ within a core radius of $4\times 10^{10}\,$cm\footnote{Note
that these cores are not completely comparable, since we assumed that
the ambient core matter was in quasi-static dynamical equilibrium and had
the same pressure and temperature distributions in all cases. This
implies slightly different gravitational accelerations 
(and hence different core masses).}. 
The results of these simulations are shown in Figures~\ref{fig:core}
and \ref{fig:core2}.

As Figure~\ref{fig:core} shows, the core--stream interaction
changes qualitatively and quantitatively, where the penetration 
depth decreases with increasing core rotation. In the case of a rapidly
rotating core, the penetration depth is centrifugally limited (i.e. by
the angular momentum in the stream). In Figure~\ref{fig:core2} we
show the logarithm of the adiabatic constant and the specific 
angular momentum near the
center of the streams in the four simulations as a function of radius.
In the two simulations with no or slow core rotation (solid and dashed
curves), the stream's angular velocity is efficiently being braked
below a radius $2.6\times 10^{10}\,$cm by its interaction with the 
core and reaches the deepest point of penetration (marked as
filled circles) where the adiabatic constant is comparable to the ambient
one. The specific angular momentum is always substantially less than
the local critical angular momentum for centrifugal support,
given by $\sqrt{gR^3}$, where $g$ is the gravitational acceleration
at radius $R$. On the other hand, in the two cases with rapid
core rotation (dot-dashed and dot-dot-dot-dashed curves), the stream
is first accelerated in the azimuthal direction by the interaction
with the core, in fact reaching an angular velocity that is larger than the
local velocity of the core (even though the initial
angular velocity at the core boundary was less). At the point of deepest
penetration, the specific angular momentum is close to the critical
angular momentum, and hence the stream becomes centrifugally supported. 
Note also that, in this case, the stream does not bounce but starts to form
a ring orbiting the core and gradually mixing with it.
This demonstrates that the main effect of rapid core rotation is to
{\em prevent/reduce} the azimuthal braking of the stream. While this leads
to less entropy generation in the stream (see the top panel in 
Fig.~\ref{fig:core2}), which ordinarily would mean that the stream should
be able to penetrate deeper, it makes the angular momentum of the stream itself
a key factor in limiting the penetration depth. In the simulation with
the fastest core, the stream matter is still substantially 
denser than the ambient matter at the final point shown in the simulations. 
Since it is also rotating faster than the local core material, 
it is reasonable to expect that the stream will continue to spiral 
in as it is being braked by the ambient medium and will penetrate somewhat
deeper than is shown in the figures. (We were unable to follow the subsequent 
evolution since the stream matter was leaving the domain of the 
calculation at this point.)

Even in the case where the core is non-rotating, the specific angular
momentum in the stream increases during a portion of its trajectory
(see the solid curve in the bottom panel between 3 and $3.8\times
10^{10}\,$cm). This is a direct consequence of the stream splitting
into two components in the impact region, a forward and a backward
component (this can be seen more clearly in Figs~1 and 2). 
As matter in the forward component is pushed by matter
following from behind, it is being accelerated and gains angular
momentum, while matter flowing backwards attains negative angular
momentum.

Finally, note that because of the narrower stream in the
post-impact region of the more rapidly rotating core, stream material
mixes more efficiently with the core material (Fig.~\ref{fig:core}
and \ref{fig:core2}). This may have been important consequences
for the nucleosynthesis in this region (see \S~5.4 and IP).

\subsection{The effects of nuclear burning}
 
\begin{figure*}
\includegraphics[scale=0.5]
 	{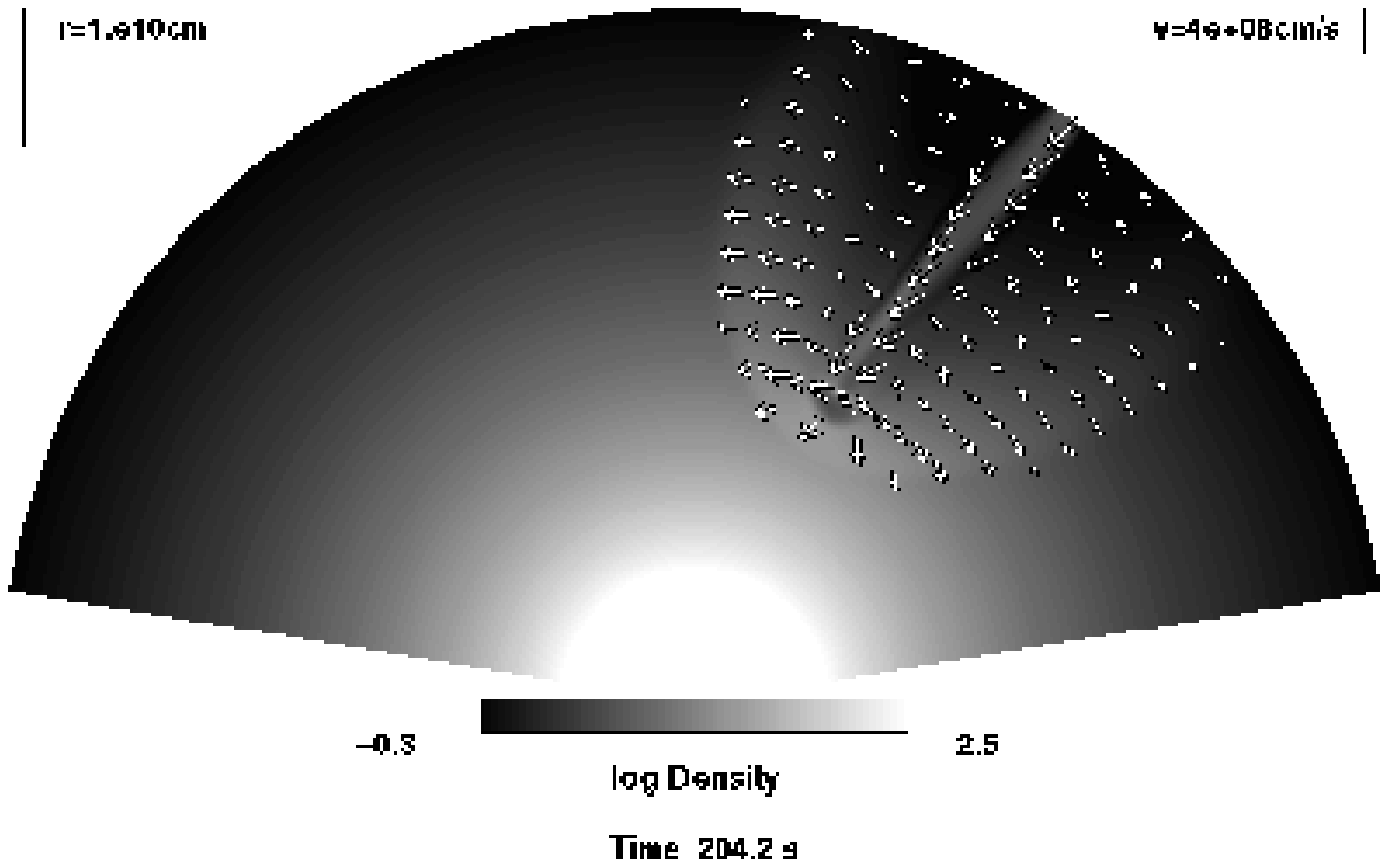}
\includegraphics[scale=0.5]
 	{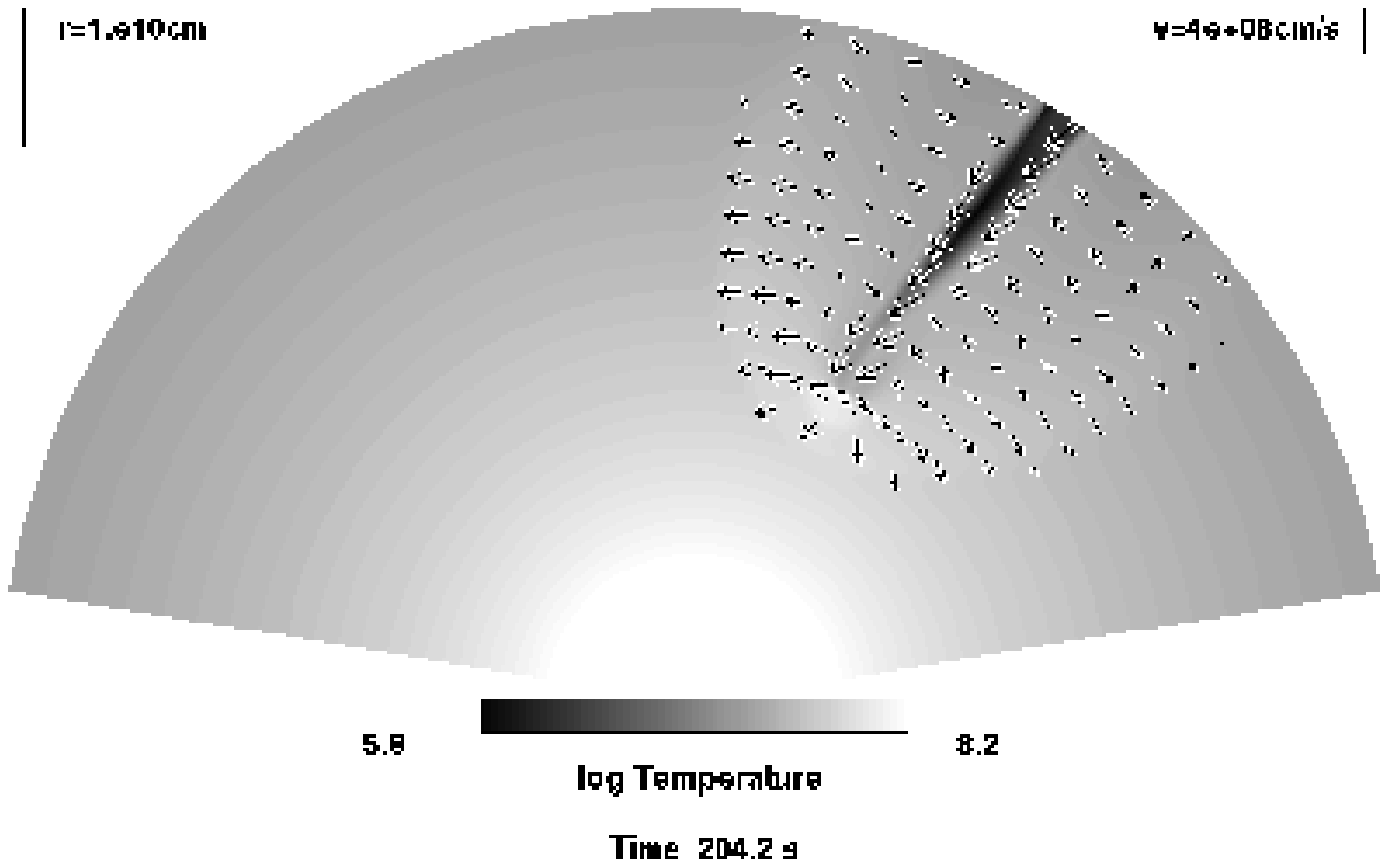} 
\includegraphics[scale=0.5]
 	{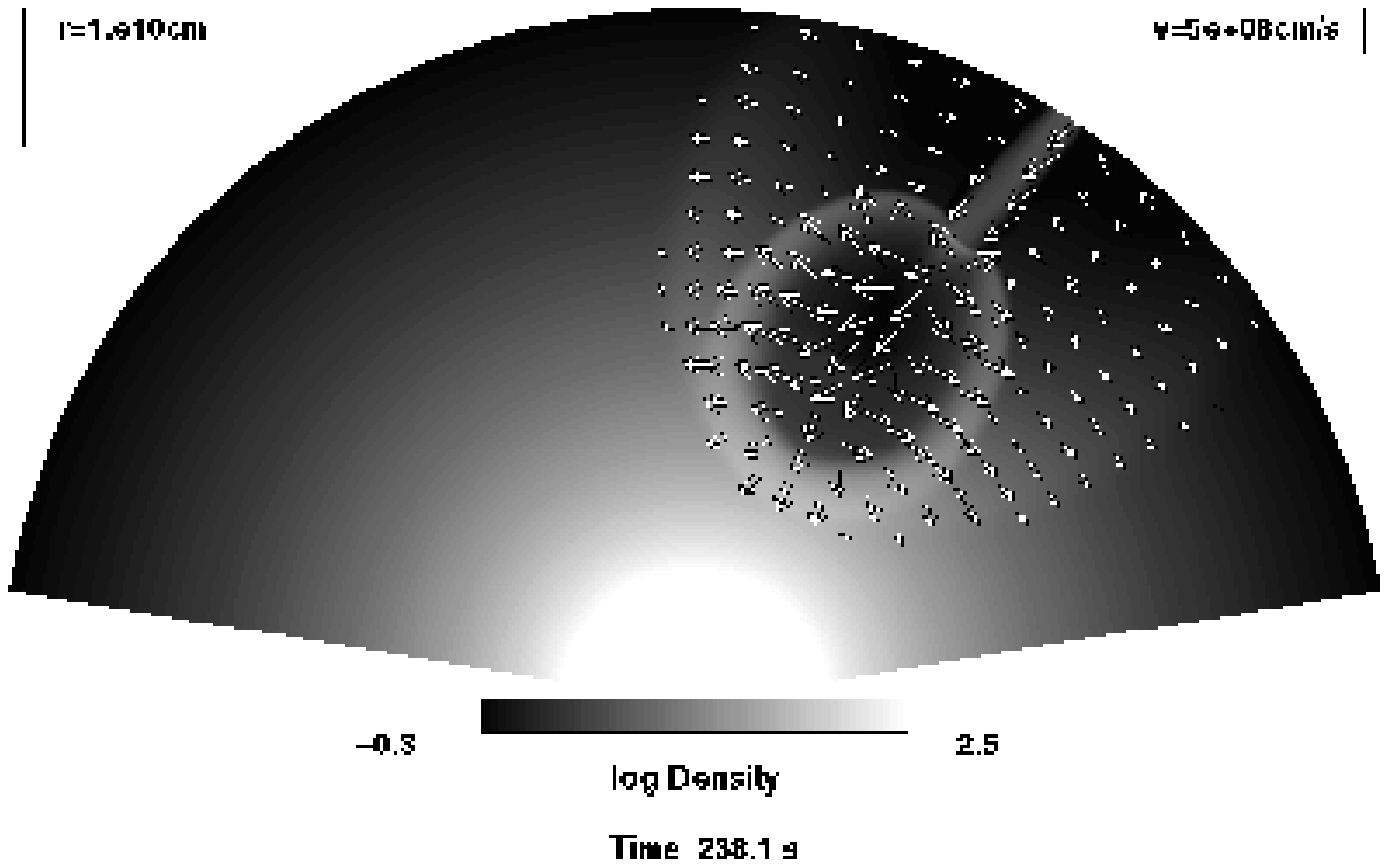}
\includegraphics[scale=0.5]
 	{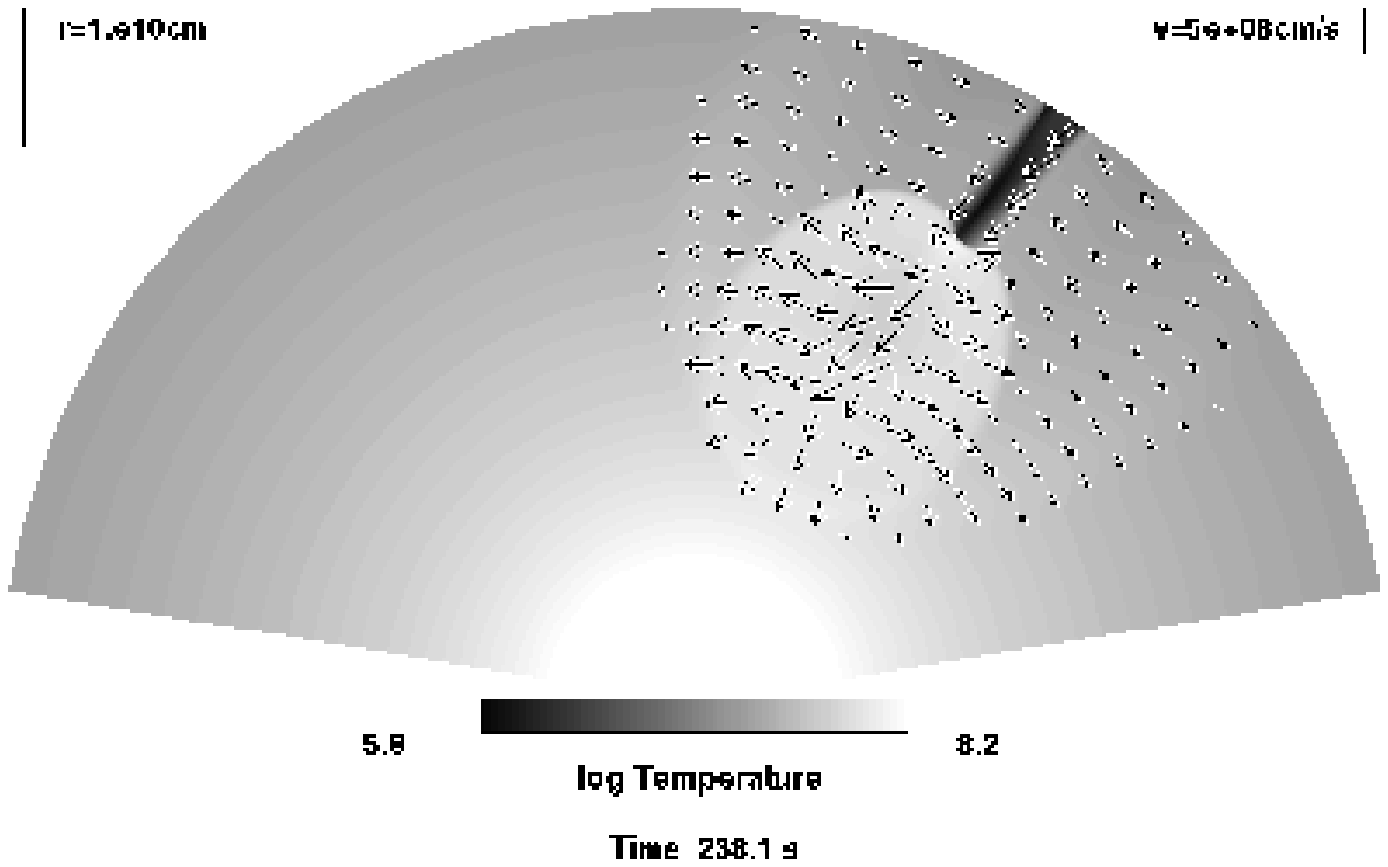} 
\includegraphics[scale=0.5]
        {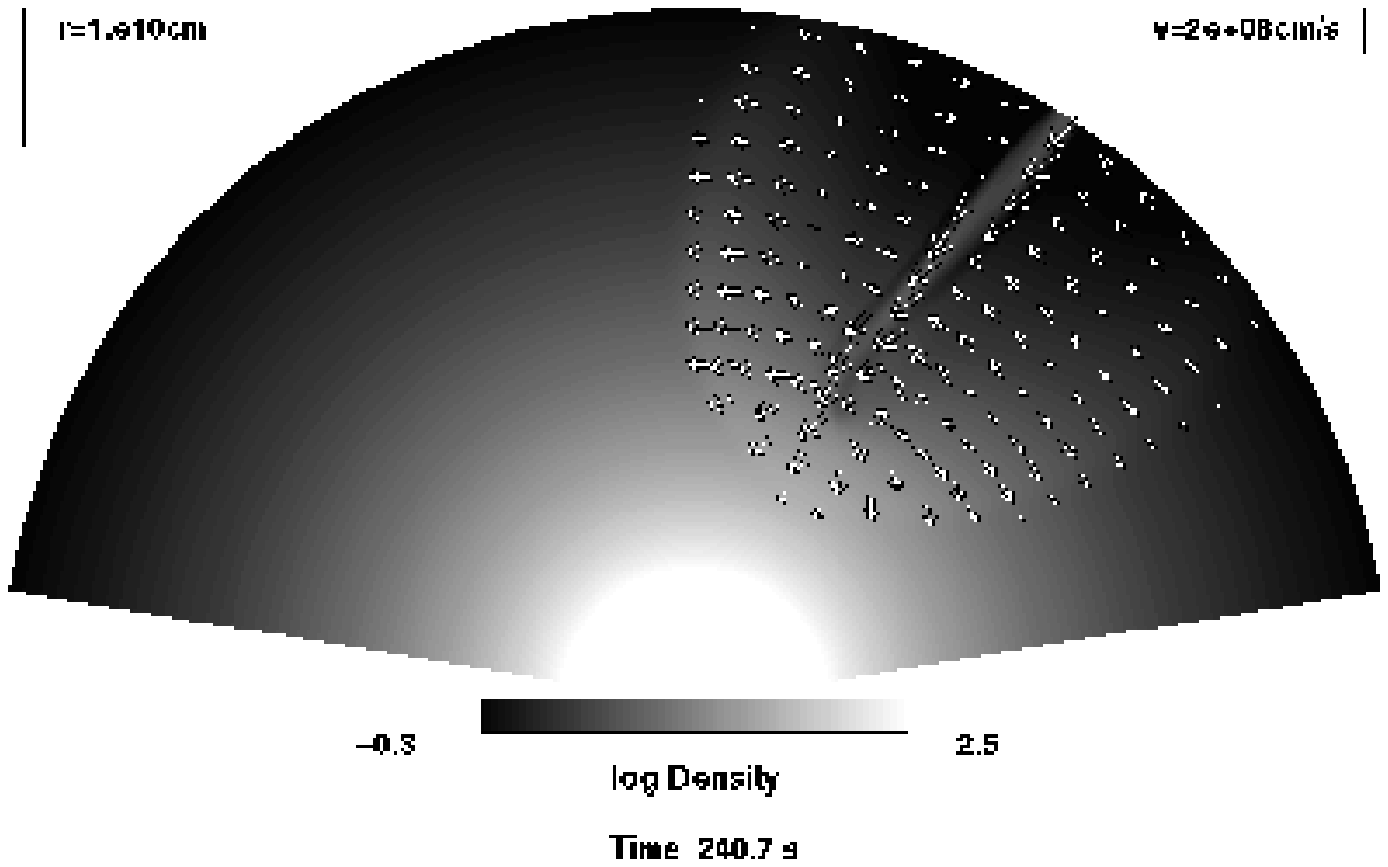}
\includegraphics[scale=0.5]
        {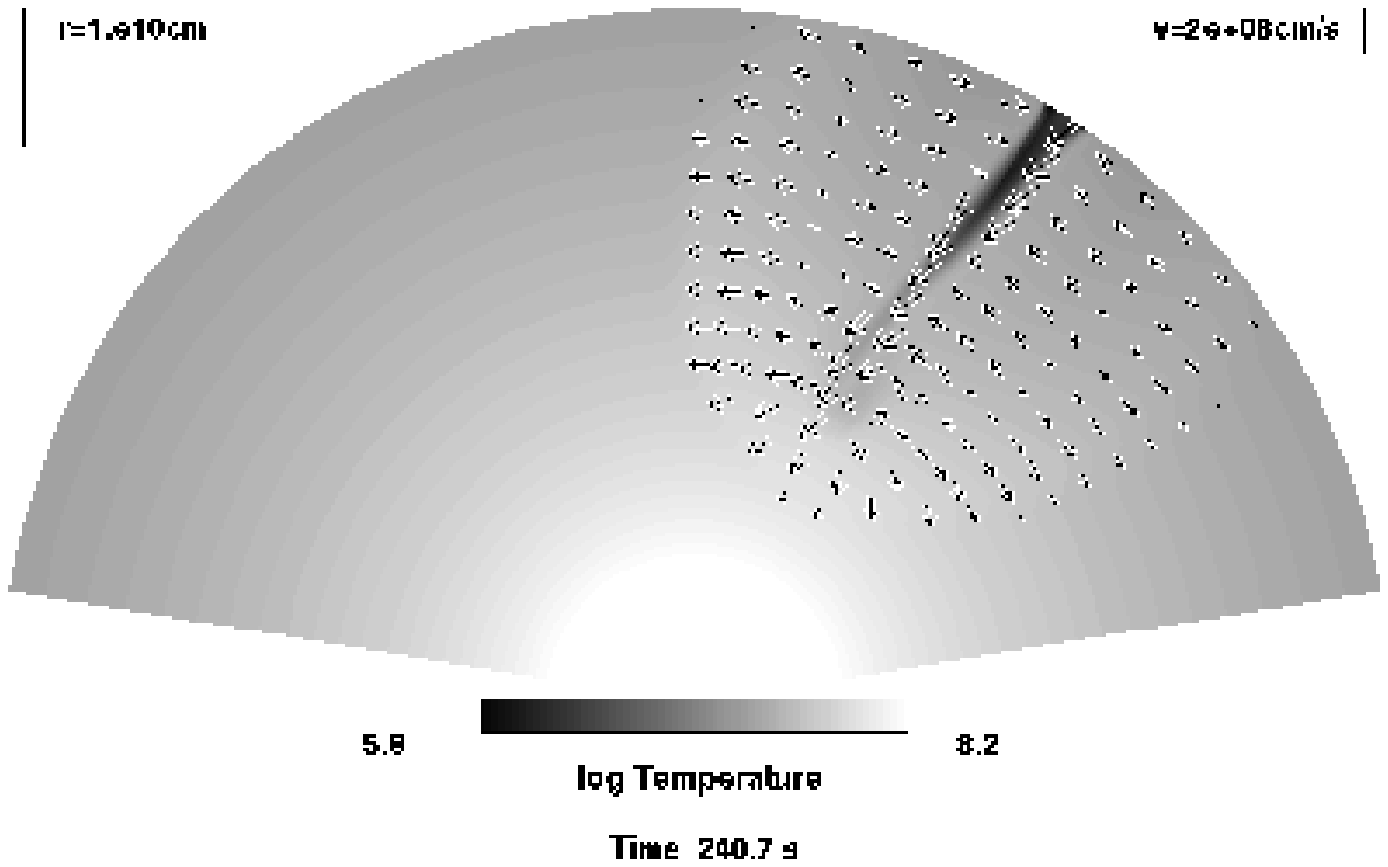}
\caption{ \label{fig:nuc} 
Density and temperature distributions for stream simulations
illustrating the effect of violent nuclear ignition. The top two panels
show the stream just before impact with the core and the point of
nuclear ignition. The middle two panels show the detonation
wave that develops shortly after nuclear ignition that disrupts the
inflowing stream. The bottom two panels show a stream at the same
time if nuclear burning is not included for comparison.
The stream parameters are
$M_{\rm int} = 9.9$, $M_{\rm ext} = 4.$, $\eta_\rho = 7$, 
$R_{\rm T } = 5\times 10^{10}\,$cm, $R_{\rm s } = 0.03 \pi$,
$\Theta=10\degr$.
}
\end{figure*}

One of the objectives of our study is to investigate the
nucleosynthesis in the stream material after it has been heated by its
impact with the core of the primary. The associated heating by
nuclear burning itself may then affect the hydrodynamics
of the stream. As can already be seen from the simulations presented
so far, the characteristic time-scale for the stream infall is of order
$1000\,$s, where the stream material typically spends no more than
$50-100\,$s near the impact region before it bounces or forms a
ring orbiting the core (in the case of a fast rotating core).  Nuclear
burning can only affect the stream motion if the characteristic
time-scale for nuclear burning is of order or shorter than the stream
dynamical time-scale.  For the largest part of the stream trajectory,
the stream material is relatively cold (for the range of stream models
considered). 

Nuclear burning only starts to increase dramatically after it has been
heated to a temperature comparable to the ambient temperature, which
occurs suddenly in the region of impact. However, since the
characteristic nuclear time-scale is generally longer than the local
dynamical time-scale, stream material will have already moved far away
from the region of impact and hence have little effect on the stream
dynamics (in the pre-impact region) and the stream trajectory. It
mainly affects the properties of the ambient matter (its composition
and temperature) and hence the background conditions for the stream on
a time-scale that is long compared to the stream infall time-scale
(these effects and the associated nucleosynthesis can be followed with
a standard stellar evolution code; see IPS for a
description of the nucleosynthesis calculations).

The situation can be different when the stream impacts the
core more-or-less radially and when the core has a very steep density 
gradient, causing a very strong shock and hence very high
post-shock temperatures.
This can cause a nuclear runaway as illustrated in Figure~\ref{fig:nuc}.
As the middle panels in this figure show, it triggers a detonation wave
that propagates up into the stream, almost completely
disrupting it in the process. The expansion wave propagating into the
ambient medium causes a dramatic expansion of the envelope material.
Once all of the matter burning in this runaway has been consumed, one
may expect that the stream re-establishes itself and that 
the process repeats, leading to pulsed nuclear burning.

\section{The penetration depth} \label{depth} 

\subsection{The entropy change}
 
In the idealized case of a stream of constant entropy, the depth 
of penetration can easily be estimated from the balancing 
of the stream ram pressure with the ambient pressure. 
In our case, the pressure of the stream material
becomes comparable to the ambient pressure well before the 
stream reaches its deepest point. In this case it is better
to use an estimate based on the conservation  
of the Bernoulli integral for matter along a streamline (though
both estimates would give similar results).
However, since the stream entropy changes because of the compression
by shock waves propagating into the stream from the sides, one has
to use a modified Bernoulli integral to estimate the penetration depth. This
is possible as long as the stream is wide enough that it is not strongly 
affected by the K-H instability and that core rotation plays a minor role.

For this purpose, we rewrite the Bernoulli integral for stream material
in the form:

\beq
\Phi_{\mbt T } + 
       {\frac  {u_{\mbt S,T }^2} {2}  } 
      +  {\frac {c_{\mbt S,T }^2}
      {\gamma_{\mbt S,T }-1} } = 
       \Phi_{\mbt B } + {\frac {u_{\mbt S,B }^2} {2} } + 
	{\frac {c_{\mbt S,B }^2}
      {\gamma_{\mbt S,B }-1}} \ ,
\eeq

\noi where $u$ is the velocity of stream material, $c$ the 
sonic speed of the stream material and $\Phi$ the potential.
The subscript S corresponds to matter in the stream,
the subscript T refers to the initial conditions and the subscript B to
the conditions at the point of deepest penetration. The point of deepest
penetration can then be determined by the condition that the
pressure of the stream is equal to the pressure in the
ambient matter and that the final stream velocity is zero (consistent
with the results of our hydrodynamical calculations).

For the  stellar structure parametrized by the pressure gradient
the corresponding potential can be written as 

\beq
\Phi(r) = -\alpha_{\mbt P } {\frac {P(r)} {\rho (r)} } = 
 -\alpha_{\mbt P } {\frac {c_{\mbt A }^2 (r) }  
{\gamma_{\mbt A } (r)} } \ ,
\eeq

\noi where the subscript $\mbox A$ corresponds to the ambient matter 
at the initial point of the stream.
Assuming that the entropy of the stream changes according to

\beq
\left ( {\frac {P} {\rho^{\gamma }} }\right) _{\mbt S,B } = 
K_{\rm s }\left ( {\frac {P} {\rho^{\gamma }} }\right) _{\mbt S,T },
\eeq

\noi one can rewrite the Bernoulli integral 
to find $K_{\rm s}$ as a function
of relative penetration depth, $\delta = r/r_{\rm T}$, and 
the initial conditions for the stream and ambient matter:

\beq
\lefteqn{ -{\frac {\alpha_{\mbt P }} {\gamma_{\mbt A }} } +
{\frac {M_{\rm ext }^2} {2} } + 
{\frac {M_{\rm ext }^2} {M_{\rm int }^2} }{\frac {1} {\gamma_{\mbt S,T } - 1 }}  = 
 -{\frac {\alpha_{\mbt P }} {\gamma_{\mbt A }} }
{\frac {\beta{\mbt T }} {\beta (\delta )}} \delta^{-\alpha_{\mbt T }} + }
\eeq
\begin{eqnarray}
{\frac {1} {\gamma_{\mbt A } }} 
{\frac {\gamma_{\mbt S,B }} { \gamma_{\mbt S,B } - 1}}
\left ( K_{\mbt s } \eta_\rho^{1-\gamma_{\mbt S,T }}
\rho_{\mbt A } ^ { -\gamma_{\mbt S,T }}
{\frac {M_{\rm ext }^2} {M_{\rm int }^2} }
{\frac {\gamma_{\mbt A }} {\gamma_{\mbt S,B } }} 
\right ) ^ {\frac {1} {\gamma_{\mbt S,B } }}
\rho_{\mbt A }
\delta^{ {\frac {1} {\gamma_{\mbt S,B } }} - \alpha_{\mbt P } }.
\nonumber
\end{eqnarray}

\noi Here $\rho_{\mbt A }$ is the density of the ambient material at the
initial radius of the steam, and $\beta$ is the ratio of gas 
pressure to total pressure for the ambient matter.

\subsection{Numerical fits}

To determine the entropy change factor, $K_{\rm s }$, we performed sets of 
hydrodynamical simulations for two parametrized structures of the ambient
medium, $(\alpha_{\rm P };\alpha_{\rm T })=(5.2.1.3)$ and $(4.2;0.9)$,
representing typical profiles in massive CE systems.
We then calculated the stream penetration depth for
3 values of density ratios, $\eta_\rho = 2$, 10 and 30, and
for a number of models with different Mach number ratios, 
$M_{\rm int} /M_{\rm ext } $ (where we varied $M_{\rm int }$, keeping
$M_{\rm ext }$ fixed).
All streams were calculated with an initial stream inclination of
$18\degr$. 

Using the expression for a weak shock (equ.~25), we fit the results
to
\beq
K_{\rm s } = 1 + k(\alpha_{\rm P },\alpha_{\rm T }) 
\cdot \eta_{\rho}^{\gamma-1} 
\left ({\frac {M_{\rm int}} {M_{\rm ext }}} \right ) ^2 \ .
\eeq

\begin{figure}
\includegraphics[scale=0.4]{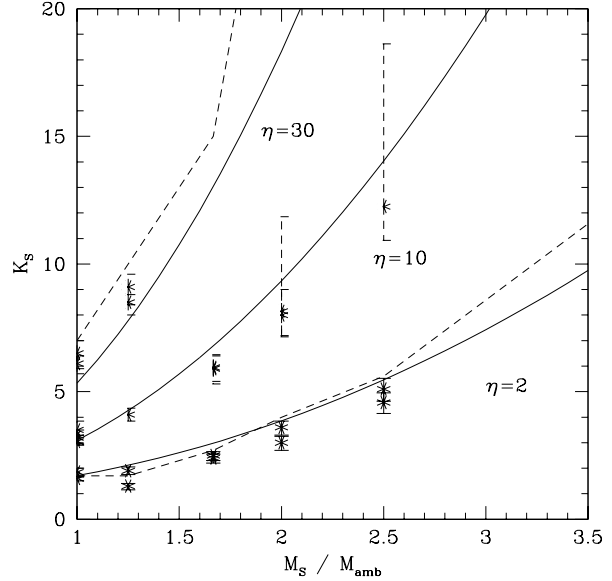}
\caption{ \label{fig:ecoef} 
The entropy change as the function of 
the Mach numbers ratio for a parametrized ambient structure with 
$\alpha_{\rm P }=5.2$ and $\alpha_{\rm T }=1.3$.
Asterisks correspond to values of $K_{\rm S } $ obtained from the
hydrodynamical simulations by determining the appropriate value
for $K_{\rm s}$ that would give the same penetration depth when using
the Bernoulli integral as in the actual hydrodynamical simulation. 
The error bars are obtained by assuming an error in the determination
of the penetration depth of 3 per cent. 
Solid curves correspond to the theoretical fit with $k =0.45$,
while the dashed curves show the change in entropy of the
stream material in the hydrodynamical simulations.}
\end{figure}

\noi For an ambient medium, characterized by $\alpha_{\rm P }=5.2;\
\alpha_{\rm T }=1.3$, these results are well fitted by a constant $k =
0.45$ and for a structure with $\alpha_{\rm P }=4.2;\ \alpha_{\rm T
}=0.9$ by $k = 0.1$ (see Fig.~\ref{fig:ecoef}).

\section{Conclusions} \label{discuss}

We have investigated the physics of the stream interaction with the
core of a massive star in the phase where a star immersed in a common
envelope merges with the core and systematically explored how the
penetration depth depends on the initial properties of the stream and
the core structure. For simulations representing the merger of a 1 and
5\Msun\ with an evolved 20\Msun\ supergiant, it typically penetrates
to a depth of $1-2\times 10^{10}\,$cm, where it is being mixed
vigorously with core material and is heated to temperatures $\ga
10^{8}\,$K. In the case of a slowly rotating core, the penetration
depth mainly depends on the initial entropy and the entropy that is
generated in the interaction. This also depends on the initial width
of the stream (which is mainly a function of the mass-transfer rate),
since narrow streams are unstable to a Kelvin-Helmholtz instability.
The main role of angular momentum in the stream is to determine the
angle of incidence of the stream (which affects the dissipation in
the stream). However, for a slowly rotating core, it is efficiently
braked by the interaction with the core, spinning up the core
in the process.  In the case of a rapidly
rotating core, the stream gains angular momentum from the core and the
penetration is centrifugally limited. The stream behaviour is
qualitatively and quantitatively different in the two cases. If the
core is rotating slowly, the stream bounces off the core, quickly
expanding away from the core, while in the case of a rapidly rotating
core, the stream merges with the core in the form of a centrifugally
supported ring.

In most cases, nuclear burning in the stream, which only becomes
effective near the region of impact, does not affect the structure and
dynamics of the stream itself, but affects the ambient medium which
affects the long-term evolution of stream--core interaction.  On the
other hand, in the case of a very strong shock (in the case of a steep
density gradient in the core and more or less normal incidence),
shock-triggered nucleosynthesis can trigger a detonation wave that
propagates upstream, temporarily disrupting the stream.

Using a simple model for the dissipation in a weak shock, we derived
a simple recipe that allows the approximate determination of the
penetration depth for a wide range of conditions. In a follow-up
paper (IP), we will implement these results in a modified stellar-evolution
code to model the whole merging phase of two stars in a slow merger.

\end{document}